\documentclass[twocolumn]{aastex63}
\pdfoutput=1

\usepackage{multirow}
\usepackage{makecell}
\usepackage{ulem}

\usepackage{balance}
\usepackage{enumerate}
\usepackage{mathrsfs}
\usepackage{graphicx}
\usepackage{subfigure}
\usepackage{booktabs}
\usepackage{hyperref}
\hypersetup{backref,pdfpagemode=FullScreen,colorlinks=true,linkcolor=blue,citecolor=blue}
\usepackage{amsmath,bm}
\usepackage{diagbox}
\usepackage{slashbox}
\usepackage[figuresright]{rotating}

\def\nh{\textit{$N_{\rm H}$}}
\def\lognh{\textit{${\rm log}\,N_{\rm H}$}}

\def\um{\textit{$\rm \mu m$}}

\def\ergs{\textit{$\rm erg\ s^{-1}$}}

\def\msun{\textit{$M_\odot$}}
\def\msunyr{\textit{$M_\odot\ \rm yr^{-1}$}}

\def\loglmirbar{\textit{$\overline{{\rm log}\,L_{\rm 6\,\um}}$}}

\def\cm{\textit{$\rm 10^{23}\ cm^{-2}$}}
\def\ct{\textit{$\rm 10^{24}\ cm^{-2}$}}
\def\>{\textit{$\textgreater$}}
\def\<{\textit{$\textless$}}
\def\hst{\textit{HST}}
\def\herschel{\textit{Herschel}}
\def\spitzer{\textit{Spitzer}}

\def\ms{\textit{$M_*$}}
\def\logms{\textit{${\rm log}\,M_*$}}
\def\logsfr{\textit{$\rm log\,SFR$}}
\def\logsfrbar{\textit{$\rm \overline{\rm log\,SFR}$}}

\def\logsfrbarm{\textit{$\rm \overline{\rm log\,SFR_{\rm 50th}}$}}
\def\logsfrbarh{\textit{$\rm \overline{\rm log\,SFR_{\rm 80th}}$}}
\def\deltalogsfrbarl{\textit{$\overline{\rm \Delta log\,SFR_{\rm 20th}}$}}
\def\deltalogsfrbarm{\textit{$\overline{\rm \Delta log\,SFR_{\rm 50th}}$}}
\def\deltalogsfrbarh{\textit{$\overline{\rm \Delta log\,SFR_{\rm 80th}}$}}

\def\lxbar{\textit{$\overline{L_{\rm X}}$}}
\def\lmirbar{\textit{$\overline{L_{\rm 6\,\um}}$}}
\def\lratiobar{\textit{$\overline{L_{\rm 6\,\um}/L_{\rm X, obs}}$}}
\def\loglratiobar{\textit{$\overline{{\rm log}\,L_{\rm 6\,\um}/L_{\rm X, obs}}$}}

\def\mbar{\textit{$\overline{M_*}$}}
\def\logmbar{\textit{$\overline{{\rm log}\,M_*}$}}

\def\zbar{\textit{$\overline{z}$}}
\def\ssfrbar{\textit{$\overline{\rm sSFR}$}}

\def\loglxbar{\textit{$\overline{{\rm log}\,L_{\rm X}}$}}
\def\lognhbar{\textit{$\overline{{\rm log}\,N_{\rm H}}$}}
\def\lx{\textit{$L_{\rm X}$}}

\def\lmir{\textit{$L_{\rm 6 \mu m}$}}

\def\loglx{\textit{${\rm log}\,L_{\rm X}$}}

\def\lognh{\textit{${\rm log}\,N_{\rm H}$}}
\def\edd{\textit{$\rm \lambda_{Edd}$}}

\def\logedd{\textit{${\rm log\,} \rm \lambda_{Edd}$}}
\def\ledd{\textit{$L_{\rm Edd}$}}
\def\eddbar{\textit{$\overline{\rm \lambda_{Edd}}$}}

\def\fmirr{\textit{$f_{24\mu m}/f_R$}}
\def\lratio{\textit{$L_{\rm 6 \mu m}/L_{\rm X, obs}$}}

\def\lxobs{\textit{$L_{\rm X, obs}$}}
\def\um{\textit{$\rm \mu m$}}

\def\lbol{\textit{$L_{\rm bol}$}}
\def\mbh{\textit{$M_{\rm BH}$}}

\def\chandra{\emph{Chandra}}
\def\spitzer{\emph{Spitzer}}

\def\m{\textit{$M_{20}$}}

\def\cm{\textit{$\rm 10^{23}\ cm^{-2}$}}
\def\ct{\textit{$\rm 10^{24}\ cm^{-2}$}}
\def\ergs{\textit{$\rm erg\ s^{-1}$}}

\def\nhu{\textit{$\rm cm^{-2}$}}

\shorttitle{Highly obscured AGNs}
\shortauthors{Li et al.}

\begin{document}

\title{Piercing through Highly Obscured and Compton-thick AGNs in the \textit{Chandra} Deep Fields. II. \\Are Highly Obscured AGNs the Missing Link in the Merger-Triggered AGN-Galaxy Coevolution Models?}

\author{Junyao Li}
\affiliation{CAS Key Laboratory for Research in Galaxies and Cosmology, Department of Astronomy, University of Science and Technology of China, Hefei 230026, China; lijunyao@mail.ustc.edu.cn, xuey@ustc.edu.cn}
\affiliation{School of Astronomy and Space Science, University of Science and Technology of China, Hefei 230026, China}

\author{Yongquan Xue}
\affiliation{CAS Key Laboratory for Research in Galaxies and Cosmology, Department of Astronomy, University of Science and Technology of China, Hefei 230026, China; lijunyao@mail.ustc.edu.cn, xuey@ustc.edu.cn}
\affiliation{School of Astronomy and Space Science, University of Science and Technology of China, Hefei 230026, China}

\author{Mouyuan Sun}
\affiliation{Department of Astronomy, Xiamen University, Xiamen, Fujian 361005, China}

\author{William N. Brandt}
\affiliation{Department of Astronomy  \& Astrophysics, 525 Davey Lab, The Pennsylvania State University, University Park, PA 16802, USA}
\affiliation{Institute for Gravitation and the Cosmos, The Pennsylvania State University, University Park, PA 16802, USA}
\affiliation{Department of Physics, The Pennsylvania State University, University Park, PA 16802, USA}

\author{Guang Yang}
\affiliation{Department of Physics and Astronomy, Texas A\&M University, College Station, TX 77843-4242, USA}
\affiliation{George P. and Cynthia Woods Mitchell Institute for Fundamental Physics and Astronomy, Texas A\&M University, College Station, TX 77843-4242, USA}
\affiliation{Department of Astronomy  \& Astrophysics, 525 Davey Lab, The Pennsylvania State University, University Park, PA 16802, USA}
\affiliation{Institute for Gravitation and the Cosmos, The Pennsylvania State University, University Park, PA 16802, USA}

\author{Fabio Vito}
\affiliation{Instituto de Astrofisica and Centro de Astroingenieria, Facultad de Fisica, Pontificia Universidad Catolica de Chile, Casilla 306, Santiago 22, Chile}
\affiliation{Chinese Academy of Sciences South America Center for Astronomy, National Astronomical Observatories, CAS, Beijing 100012, China}

\author{Paolo Tozzi}
\affiliation{Istituto Nazionale di Astrofisica (INAF) -- Osservatorio Astrofisico di Arcetri, Largo Enrico Fermi 5, I-50125 Firenze Italy}

\author{Cristian Vignali}
\affiliation{Dipartimento di Fisica e Astronomia, Alma Mater Studiorum, Universit\`a degli Studi di Bologna, Via Gobetti 93/2, I-40129 Bologna, 
Italy}
\affiliation{INAF -- Osservatorio di Astrofisica e Scienza dello Spazio di Bologna, 
Via Gobetti 93/3, I-40129 Bologna, Italy}

\author{Andrea Comastri}
\affiliation{INAF -- Osservatorio di Astrofisica e Scienza dello Spazio di Bologna, 
Via Gobetti 93/3, I-40129 Bologna, Italy}

\author{Xinwen Shu}
\affiliation{Department of Physics, Anhui Normal University, Wuhu, Anhui, 241000, China}

\author{Guanwen Fang}
\affiliation{Institute for Astronomy and History of Science and Technology, Dali University, Dali 671003}

\author{Lulu Fan}
\affiliation{CAS Key Laboratory for Research in Galaxies and Cosmology, Department of Astronomy, University of Science and Technology of China, Hefei 230026, China; lijunyao@mail.ustc.edu.cn, xuey@ustc.edu.cn}
\affiliation{School of Astronomy and Space Science, University of Science and Technology of China, Hefei 230026, China}

\author{Bin Luo}
\affiliation{School of Astronomy and Space Science, Nanjing University, Nanjing 210093, China}
\affiliation{Key Laboratory of Modern Astronomy and Astrophysics (Nanjing University), Ministry of Education, Nanjing, Jiangsu 210093, China}
\affiliation{Collaborative Innovation Center of Modern Astronomy and Space Exploration, Nanjing 210093, China}

\author{Chien-Ting Chen}
\affiliation{Astrophysics Office, NASA Marshall Space Flight Center, ZP12, Huntsville, AL 35812}

\author{Xuechen Zheng}
\affiliation{Leiden Observatory, Leiden University, PO Box 9513, NL-2300 RA Leiden, the Netherlands}

\begin{abstract}
\noindent By using a large highly obscured ($\nh > \cm$) AGN sample (294 sources at $z \sim 0 - 5$) selected from detailed X-ray spectral analyses  in the deepest \chandra~surveys, we explore distributions of these X-ray sources in various optical/IR/X-ray color-color diagrams and their host-galaxy properties, aiming at characterizing the nuclear obscuration environment and the triggering mechanism of highly obscured AGNs.
We find that the refined IRAC color-color diagram fails to identify the majority of X-ray selected highly obscured AGNs, even for the most luminous sources with $\loglx\, (\ergs) >44$. Over 80\% of our sources will not be selected as heavily obscured candidates using the flux ratio of $\fmirr > 1000$ and $R-K > 4.5$ criteria, implying complex origins and conditions for the obscuring materials that are responsible for the heavy X-ray obscuration. 
The average star formation rate of highly obscured AGNs is similar to that of stellar mass- ($M_*$-) and $z$-controlled normal galaxies, while the lack of quiescent hosts is observed for the former.
Partial correlation analyses imply that highly obscured AGN activity (traced by \lx) appears to be more fundamentally related to $M_*$, and no dependence of \nh~on either $M_*$ or SFR is detected.
Morphology analyses reveal that 61\% of our sources have a significant disk component, while only $\sim 27\%$ of them exhibit irregular morphological signatures. These findings together point toward a scenario where secular processes (e.g., galactic-disk instabilities), instead of mergers, are most probable to be the leading mechanism that triggers accretion activities of X-ray-selected highly obscured AGNs.
\end{abstract}


\section{Introduction} \label{sec:intro}
Since the observational establishment that there are tight correlations between the masses of supermassive black holes (SMBHs) and their host-galaxy properties (such as stellar velocity dispersion) in the local universe, how such small-scale SMBHs coevolve with their large-scale host galaxies has become one of the most fundamental problems in understanding the evolution of galaxies (see, e.g., \citealt{Kormendy2013} for a review).  
Merger-triggered coevolution models \citep[e.g.,][]{Sanders1988, Dimatteo2005, Hopkins2006}, in which the gas-rich major merger induces both intense star formation and obscured active galactic nucleus (AGN) activity while the subsequent AGN feedback eventually sweeps out the obscuring materials and shuts down the growth of both the SMBH and stellar populations, provide an attractive explanation to how the central AGN communicates with and influences its host galaxy. 

Many studies have been devoted to searching for the possible connections between AGN luminosity,  obscuration and host-galaxy properties, such as stellar mass ($M_*$), star formation rate (SFR) and merger signatures, to test the merger-driven evolutionary models \cite[e.g.,][]{Lutz2010, Mainieri2011, Schawinski2012, Chen2013, Stanley2015, Donley2018}. 
However, how AGN activities are triggered and the exact role that mergers/AGNs play in regulating SMBH/galaxy growth are still under debate. The merger fractions are found to be generally low in various AGN populations \citep[typically $\lesssim 20\%;$ e.g.,][]{Silverman2011, Kocevski2012, Schawinski2012, Villforth2014, Lackner2014, Hewlett2017}, even for those obscured quasars \citep[e.g.,][]{Zhao2019} or fast-accreting AGNs \citep[e.g.,][]{Villforth2017, Marian2019} where we may expect to see a higher incidence of merger signatures (but see \citealt{Treister2012}). 
A positive correlation between galaxy-wide star formation and AGN activities has been reported in several works, at least for the luminous populations \citep[e.g.,][]{Lutz2010, Shao2010, Hatziminaoglou2010, Rovilos2012, Rosario2012, Chen2013, Dai2018}, but others find a flat relationship \citep[e.g.,][]{Stanley2015, Suh2017, Schulze2019} or suggest that SMBH accretion is probably linked to a complex combination of galaxy properties including \ms, SFR and morphology \citep[e.g.,][]{Rodighiero2015, Yang2017, Fornasini2018, Yang2019, Ni2019}, especially that the time-averaged black hole accretion rate (BHAR) appears to be only correlated with bulge growth \citep{Yang2019}. The suppression of star formation at high AGN luminosities has been reported only in a few works \citep[e.g.,][]{Page2012, Barger2015}, while \cite{Harrison2012} pointed out that such observed negative AGN feedback may be simply caused by low source number statistics. 

Moreover, the analyses of the link between AGN obscuration and host-galaxy properties has also presented mixed results. 
While \cite{Lanzuisi2017} claimed that the hydrogen column density (\nh) is strongly connected with $M_*$ but not SFR (also see \citealt{Rodighiero2015}), \cite{Lutz2010} and \cite{Chen2015} suggested a possible correlation between obscuration and SFR indicators. Other studies found no correlation between AGN obscuration and host properties \citep[e.g.,][]{Shao2010, Rosario2012}. 

Several factors may be responsible for the contradictory results (see, e.g., Section~3.1 of \citealt{Xue2017}; and also Section 5~of \citealt{BA2015}), including the limited sample size \citep[e.g.,][]{Harrison2012}, the different sample-selection methods (e.g., X-ray vs. IR), the adoption of different indicators to trace AGN (e.g., hardness ratio vs. \nh) and galaxy properties, how the undetected sources are treated via stacking \citep[e.g.,][]{Mullaney2015}, whether the AGN contamination is properly removed through decomposition when calculating the star formation luminosity \citep[especially when performing stacking analyses; e.g.,][]{Lutz2010, Rosario2012, Ramasawmy2019}, as well as the influence of AGN variability and the usage of different binning strategies while analyzing the correlation between two parameters which vary on different timescales \citep[e.g.,][]{Neistein2014, Hickox2014, Volonteri2015, Lanzuisi2017}. 

Furthermore, the lack of correlation between AGN and host-galaxy properties may arise because we are looking at the ``inappropriate'' AGN populations \citep[e.g.,][]{Kocevski2015, Donley2018}. Cosmological simulations suggest that most of the SMBH growth is expected to happen during a phase of heavy obscuration \citep[e.g.,][]{Hopkins2006, Hopkins2008}, traced by high \nh\ values in the X-ray band. Therefore, highly obscured AGNs (i.e., having $\nh > \cm$), which are predicted to represent a critical phase in coevolution models where the heavily dust-enshrouded environment, the enhanced star formation activity and active SMBH accretion all happen ``together'' via mergers \citep[e.g.,][]{Springel2005}, may be the ``right'' AGN population to examine such evolutionary models. 

Indeed, some studies have found that the X-ray-selected most heavily obscured Compton-thick (CT; defined as $N_{\rm H}\ge 10^{24}$~cm$^{-2}$) AGNs exhibit enhanced merger signatures relative to less-obscured AGNs \citep[e.g.,][]{Kocevski2015, Koss2016, Lanzuisi2018}. However, \cite{Schawinski2012} found that 90\% of their heavily obscured quasar candidates are hosted in disk galaxies without showing any disturbed signatures, conflicting with other studies.

In addition, the total merger fractions for X-ray-selected highly obscured AGN samples \citep[$\approx 20$\%--30\%; e.g.,][]{Kocevski2015} are found to be significantly lower than that for IR-selected luminous quasars \citep[$\approx 60$\%--80\%; e.g., ][]{ Fan2016b, Donley2018}, and their star formation activities  \citep[e.g.,][]{Georgantopoulos2013, Lanzuisi2015} also seem to be more silent than IR-selected dust-obscured AGNs \citep[e.g.,][]{Fan2016a},  further raising questions about whether highly obscured AGNs selected from various diagnostics are triggered by different mechanisms or situate in different evolutionary phases. 

In this study, we focus on the X-ray-selected highly obscured AGNs, which present the cleanest sample compared to other selection methods \citep[e.g.,][]{BA2015, Xue2017}, and ensure the most direct measurements of AGN activity (X-ray luminosity; \lx) and obscuration (\nh).
By systematically analyzing the multiwavelength data for a large dedicated X-ray-selected highly obscured AGN sample \citep[][hereafter paper I]{Li2019a} in the deepest \emph{Chandra} Deep Fields surveys \citep[CDFs; for a review, see][]{Xue2017},
we aim at comprehensively exploring (1) the AGN obscuration properties; (2) whether the growth of highly obscured AGNs is isolated in a small nuclear region or somehow linked with host galaxies; (3) the role of merger in igniting highly obscured SMBH accretion; and (4) whether such AGNs are experiencing a blow-out phase which may eventually make themselves evolve to unobscured AGNs, in order to examine whether highly obscured AGNs are the missing-link in the merger-triggered SMBH-galaxy coevolution models.

This paper is organized as follows. In $\S$ \ref{sec:data} we describe our X-ray-selected highly obscured AGN sample and the compilation of the multiwavelength data to construct their broadband spectral energy distributions (SEDs). In $\S$ \ref{sec:method} we describe our SED-fitting method to derive AGN and galaxy properties. In $\S$ \ref{sec:IR} we present the distributions of our X-ray AGNs on various optical/IR/X-ray color-color diagrams and their implications for AGN obscuration. 
In $\S$ \ref{sec:missing_link} we discuss the analyses of star formation activity of our AGN hosts, the connections between AGN properties and their host-galaxy growth, the role that mergers play in triggering highly obscured SMBH accretion, and whether highly obscured AGNs are sweeping out the surrounding materials. In $\S$ \ref{sec:conclusions} we summarize the primary conclusions emerging from this work.
Throughout this paper, we adopt flat cosmological parameters with $\rm H_0 = 70.0\ km\ s^{-1}\ Mpc^{-1}$, $\rm \Omega_M = 0.30$, and $\rm \Omega_\Lambda = 0.70$.  We define Compton-thin (CN) AGNs as having $\nh < \ct$, and AGNs with $\cm < \nh < \ct$ are called highly obscured CN AGNs. The remaining AGNs with $\nh < \cm$ are referred to as less-obscured AGNs.

\section{Multiwavelength Data}\label{sec:data}
One of the primary goals of this work is to 
characterize the host-galaxy properties of highly obscured AGNs. The most commonly used method to derive galaxy parameters, such as $M_*$ and SFR, is through fitting their SEDs. Among extragalactic surveys, CDFs are among the most extensively investigated fields which enable us to gather a wealth of multiwavelength data from the ultraviolet (UV) to far-infrared (FIR) regimes and compile broadband SEDs for sources of interest \cite[e.g.,][]{Gao2019, Guo2020}. Here we describe the multiwavelength data sets for our sample. 

\subsection{X-ray Data}
In paper I, we systematically analyzed the X-ray spectral and variability properties of a sample of 436 highly obscured AGNs (including 102 CT AGN candidates) selected in the 7~Ms CDF-S \citep{Luo2017} and 2~Ms CDF-N \citep{Xue2016} surveys, which are the two deepest \chandra~surveys to date. The mean redshift for this sample is 1.88 with 191 sources having spectroscopic redshifts and 245 sources having high-quality photometric redshifts (see Sections 2 and 4.4 of paper I). We performed detailed X-ray spectral modeling and obtained crucial AGN properties such as \nh, the observed (\lxobs) and intrinsic (\lx) $\rm 2-10\ keV$ luminosities and fluxes in the rest frame. All the relevant AGN X-ray information is taken from paper I and we refer the readers to paper~I for details of X-ray spectral fitting.

For comparison purpose, we also include 492 less obscured AGNs with $\loglx > 42\ \ergs$ identified in paper~I in our analyses. Note that the X-ray spectral fitting model we used in paper~I (i.e., MYTorus; \citealt{Murphy2009}) does not allow \nh~to vary below $10^{22}\ \nhu$. To derive a column density value for those X-ray unobscured sources, we refit their X-ray spectra by replacing the absorption (MYTZ), reflection (MYTS) and emission line (MYTL) models of MYTorus with the commonly adopted  $wabs$, $pexrav$ and $gauss$ models.  The new spectral fitting results are consistent with the previous MYTorus-based results on the classification of less obscured and highly obscured AGNs. For sources with $\nh < 10^{20}\ \nhu$, we set their \nh~values to $10^{20}\ \nhu$. 

Note that the depths and sky coverages of multiwavelength surveys significantly drop at the outskirts of the CDF-S and CDF-N.  Therefore, we restrict our analyses to the 294 highly obscured AGNs and 250 less obscured AGNs that lie within the central GOODS-S and GOODS-N fields to ensure reliable SED fitting results (see Figure \ref{fig:sample}, where the distributions of $z$, \lx~and \nh~of our sample are also shown). 
 
\subsection{UV and Optical Data}
The UV data are taken from the \textit{GALEX} DR6 catalog.\footnote{\url{http://galex.stsci.edu/GR6/}}
For the CDF-S, our optical data include $U$-, $B$-, $V$-, $R$-, $I$-, and $Z$-band photometry from the MUSYC survey \citep{Gawiser2006}; the $F606W$ and $F814W$ photometry of the \emph{Hubble Space Telescope} (\hst) from the CANDELS/3D-\hst~catalog \citep{Skelton2014}; and the $F435W$, $F775W$ and $F850LP$ data from the CANDELS multiwavelength catalogs \citep{Guo2013}. We also supplement these data with 18-band Subaru narrow-band photometry compiled in \cite{Hsu2014}. The optical data in the CDF-N are mainly from \cite{Yang2014} which collected images from \cite{Capak2004} and \cite{Ouchi2009} and presented point spread function-matched photometry in the $U$, $B$, $V$, $R$, $I$, and $z'$ bands in the H-HDF-N. The \hst~$F435W$, $F775W$ and  $F850LP$ data are adopted from the GOODS v2.0 catalog \citep{Giavalisco2004}.

\subsection{NIR and MIR Data}
We combine \hst~and \emph{Spitzer} data with ground-based near-infrared (NIR) photometry to construct the NIR to mid-infrared (MIR) SEDs. 
For the CDF-S, the $F098M$, $F105W$, $F125W$, $F140W$ and $F160W$ data are collected from \cite{Skelton2014} and \cite{Guo2013}.  The \spitzer~IRAC 3.6 \um, 4.5 \um, 5.8 \um, 8.0 \um, MIPS 24 \um~and 70 \um~data are adopted from the SIMPLE survey \citep{Damen2011} and the GOODS-\herschel~catalog \citep{Elbaz2011}. We also utilize $J_1$-, $J_2$-, $J_3$-, $H_s$-, $H_l$-, $K$- and deep $K_s$-band photometry from the ZFOURGE catalog \citep{Straatman2016}.
For the CDF-N, the \spitzer~IRAC photometry as well as the $J$-, $H$-, $K_s$- and $H_k$-band data are taken from \cite{Yang2014}. The detailed description of these data can be found in Table 1 of \cite{Yang2014}. The $F125W$, $F140W$, $F160W$ data are gathered from \cite{Skelton2014}. The \spitzer~IRS 16~\um~and 24~\um~data are taken from \cite{LiuD2018}.

\subsection{FIR Data}
The CDFs had been observed by the PACS and SPIRE instruments aboard the \textit{Herschel Space Observatory} at FIR wavelengths of 100 \um, 160 \um, 250 \um, 350 \um~and 500 \um. For the CDF-S, we combine the GOODS-\herschel~survey \citep{Elbaz2011} and the HerMES survey \citep{Oliver2012} to obtain FIR data which are calculated by adopting the \emph{Spitzer} MIPS 24 \um~positions as priors. For the CDF-N, we use the state-of-the-art ``Super-deblended'' FIR and sub-millimeter (SCUBA 850 \um~data from the James Clerk Maxwell Telescope) photometry presented in \cite{LiuD2018}. This advanced super-deblend technique can significantly improve the accuracy of the measured photometry for confused sources.

Following \cite{Stanley2015}, for part of the FIR non-detected sources that do not have flux upper limits provided by the catalogs, we use the 100~\um~and 160~\um~residual maps to derive them\footnote{\url{http://www.mpe.mpg.de/ir/Research/PEP/DR1}}. For each non-detected source, we randomly extract 1000 aperture photometry measurements in the source-free vicinity ($100''$) of the source optical position, and calculate the 99.7th percentile of the measured flux distribution as the $3\sigma$ upper limit value (see Section 4.3 of \citealt{Boquien2019} for how CIGALE handles upper limits).

\subsection{Construction of Broadband SEDs}

\begin{figure*}
\centering
\includegraphics[width=0.497\linewidth]{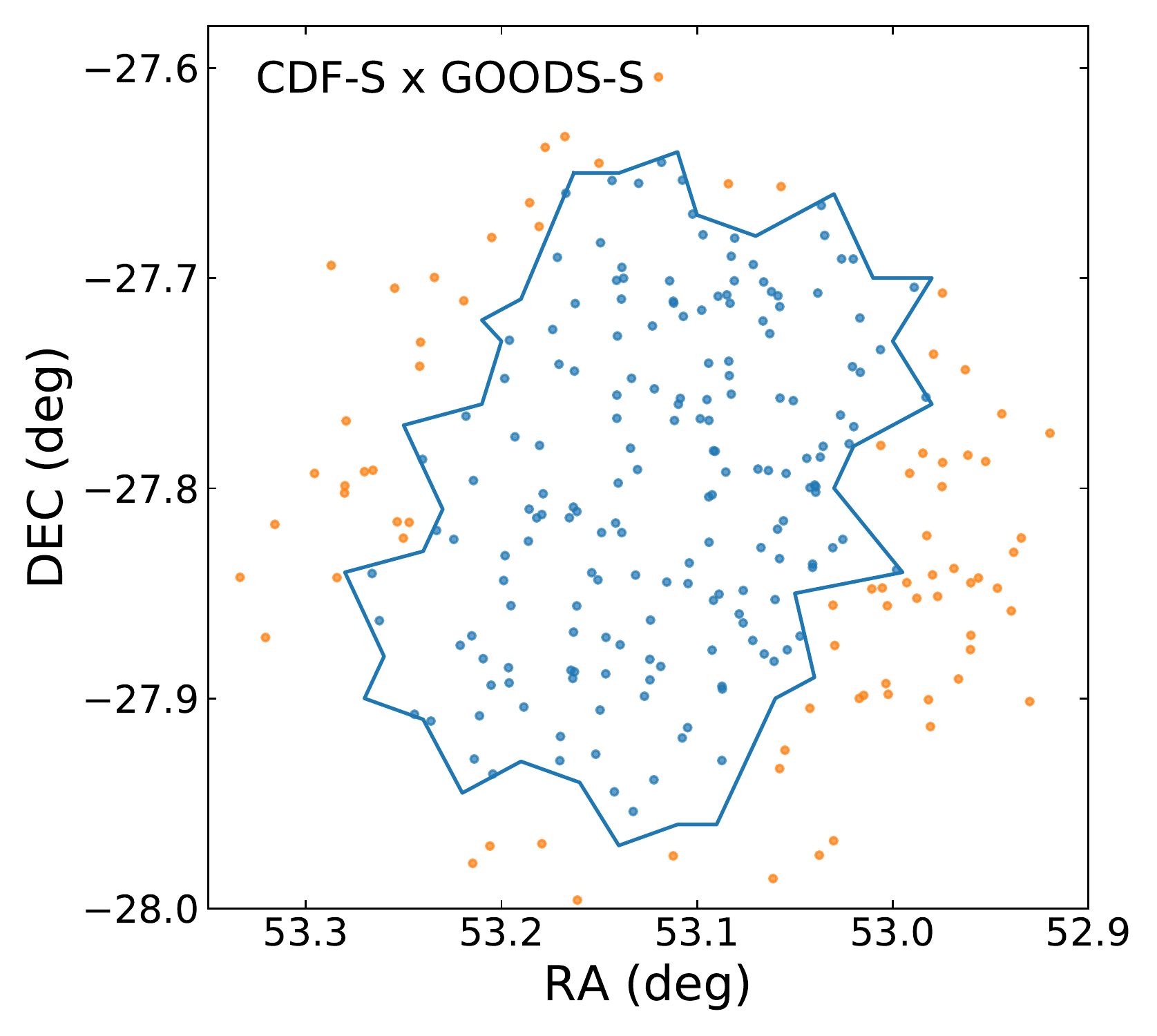}
\includegraphics[width=0.497\linewidth]{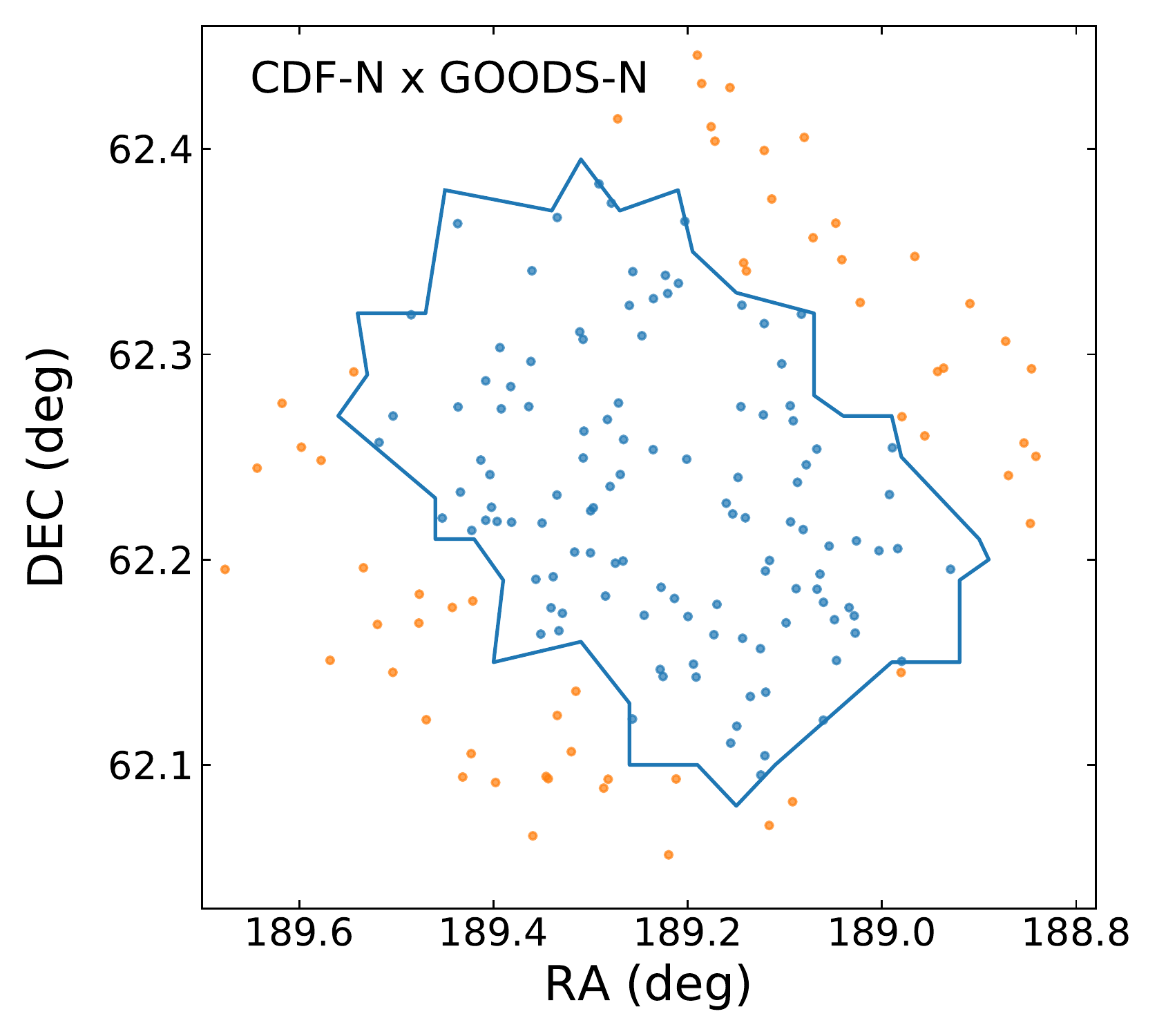}
\includegraphics[width=\linewidth]{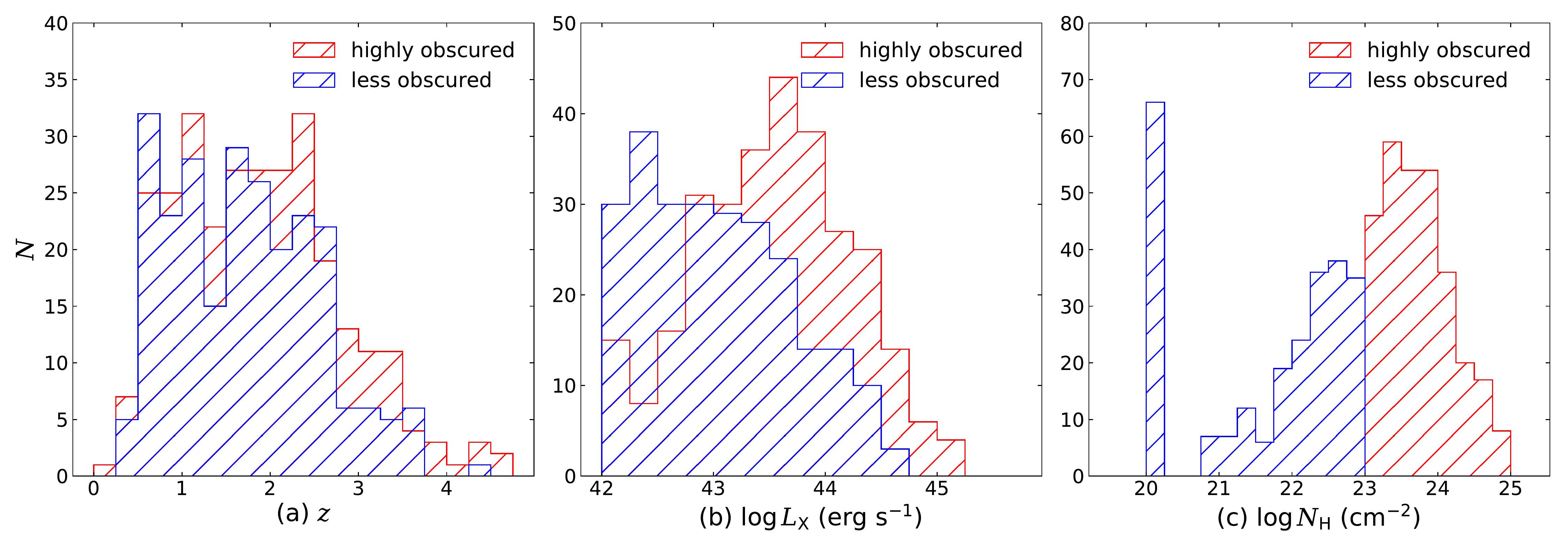}

\caption{Top: Highly obscured AGN sample (294 sources) within the GOODS fields (blue region) used in this work (blue points). Sources in paper~I that lie outside the GOODS fields are excluded (orange points). Bottom: Distributions of $z$, \lx~and \nh~for the highly obscured (red) and less obscured (blue) AGN samples used in this work.}
\label{fig:sample}
\end{figure*} 

For the optical, NIR and MIR-FIR catalogs, we adopt $0.5''$, $1''$ and $2''$ as the matching radii to cross-match with our X-ray sources using the coordinates of their multiwavelength counterparts (mostly optical ones) provided by the X-ray catalogs, respectively, and combine the matched multiwavelength data to construct the broadband SEDs. We adopt larger matching radii for IR catalogs due to the lower spatial resolution of IR images. When multiple associations are found within the matching radius, we adopt the closest one as the counterpart. The intrinsic (i.e., absorption-corrected) rest-frame 2-10~keV flux is used to represent the X-ray SED.
Among the total sample, 164 sources have at least one solid $>3\sigma$ detection in the aforementioned five \herschel~bands. For the remaining sources that lack  \herschel~detections, their SFRs may not be well constrained \citep[e.g.,][]{Gao2019}. We will discuss the influence of this issue in Section \ref{sec:method}.

\newcommand{\tabincell}[2]{\begin{tabular}{@{}#1@{}}#2\end{tabular}}

\begin{table*}
\renewcommand\arraystretch{1.0}
\footnotesize
\caption{SED fitting models and parameter spaces adopted in X-CIGALE.}
\begin{tabular}{p{9cm} c}
\hline
Parameter & Value\\
\hline
\multicolumn{2}{c}{{Stellar population synthesis model}: \cite{BC03}}\\
Initial mass function & Chabrier\\
star formation history & Delayed $\tau$ model\\
E-folding time of the main stellar population model in Myr & 100, 158, 251, 398, 631, 1000, 1584, 2512, 3981, 6309, 10000\\
Age of the oldest stars in the galaxy in Myr & 100, 158, 251, 398, 631, 1000, 1584, 2512, 3981, 6309, 10000\\
Metallicity & 0.02\\
\hline
\multicolumn{2}{c}{{Galactic dust attenuation}: \cite{Calzetti2000}}\\
E($B-V$) lines & 0.01, 0.1, 0.2, 0.3, 0.4, 0.5, 0.6, 0.7, 0.8, 0.9, 1.0\\
\hline
\multicolumn{2}{c}{{Galactic dust emission}: \cite{Dale2014}}\\
Power-law slope $\alpha$ & 1.5, 2.0, 2.5\\
\hline
\multicolumn{2}{c}{{Torus model}: SKIRTOR \citep{Stalevski2012, Yang2020}}\\
Average edge-on optical depth at 9.7 \um & 7.0\\
Angle between equatorial axis and line of sight & 30, 70\\
Half-opening angle of the torus & 40\\
AGN fraction ($agn\_frac$) & 0.01, 0.05, 0.1, 0.15, 0.2, 0.25, 0.3, 0.35, 0.4\\
& 0.45, 0.5, 0.6, 0.7, 0.8, 0.9, 0.99\\
\hline
\multicolumn{2}{c}{{X-ray model}}\\
Photon index & 1.8\\
Maximum $\Delta \alpha_{\rm ox}$ & 0.2\\
\hline
\hline
\end{tabular}

\vspace{10.0 pt}
{{\bf Note}. See \cite{Boquien2019} and \cite{Yang2020} for model details. We adopt the default values in X-CIGALE for parameters that are not listed in this table. One thing to mention is that, given that our sources are robustly selected as X-ray AGNs with $\lx > 10^{42}\ \ergs$ (see paper I), we therefore artificially require an AGN component during SED fitting by prohibiting the AGN fraction parameter (defined as the AGN contribution to the total IR luminosity) from being zero, which would allow us to measure a MIR luminosity for each AGN instead of having a zero value. The relevant results in Sections \ref{subsec:f24} and \ref{subsec:est_nh} are not affected by this choice, as utilizing 271 out of 294 sources that have best-fit $agn\_frac>0$ when allowing it to take a value of zero would yield the same conclusions. The influence of this forced lower-limit AGN contribution (i.e., 1\%) to the total IR luminosity (thus SFR) is also subtle and does not materially affect our SFR-related analyses.}

\label{table:sed_para}
\end{table*}

\begin{figure}
\centering
\includegraphics[width=\linewidth]{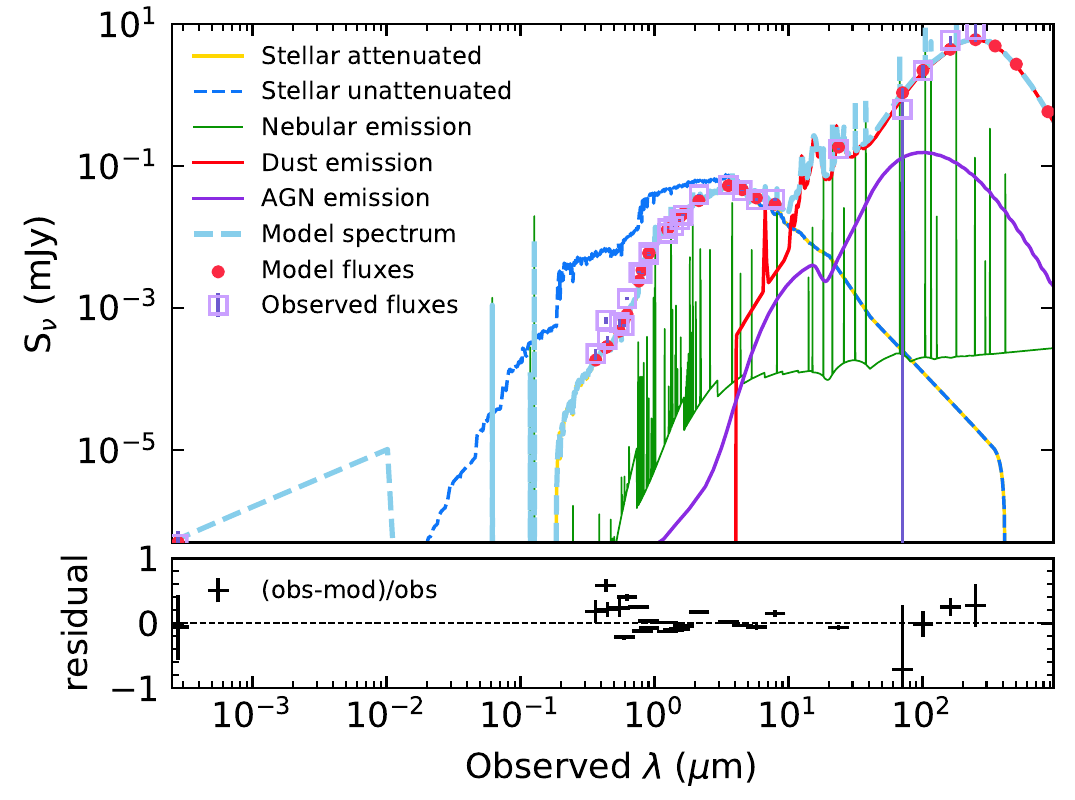}
\caption{Best-fitting SED for CDF-N XID 511 at $z=1.02$ (top panel) and fitting residuals (bottom panel), defined as (data-model)/data.Different model components are labeled by different colors. Note that the attenuated stellar emission is largely overlapped with the model spectrum 
at UV-optical wavelengths and the unattenuated stellar emission at IR wavelengths. The observed fluxes are shown in open squares while the predicted model fluxes at filter wavelengths are shown as filled circles. The SFR,  stellar mass, E(B-V) and AGN fraction for this source are 7.3~\msunyr, $10^{10.9}$~\msun, 0.9 and 5\%, respectively.}
\label{fig:sed_example}
\end{figure} 

\begin{figure}
\centering
\includegraphics[width=1.0\linewidth]{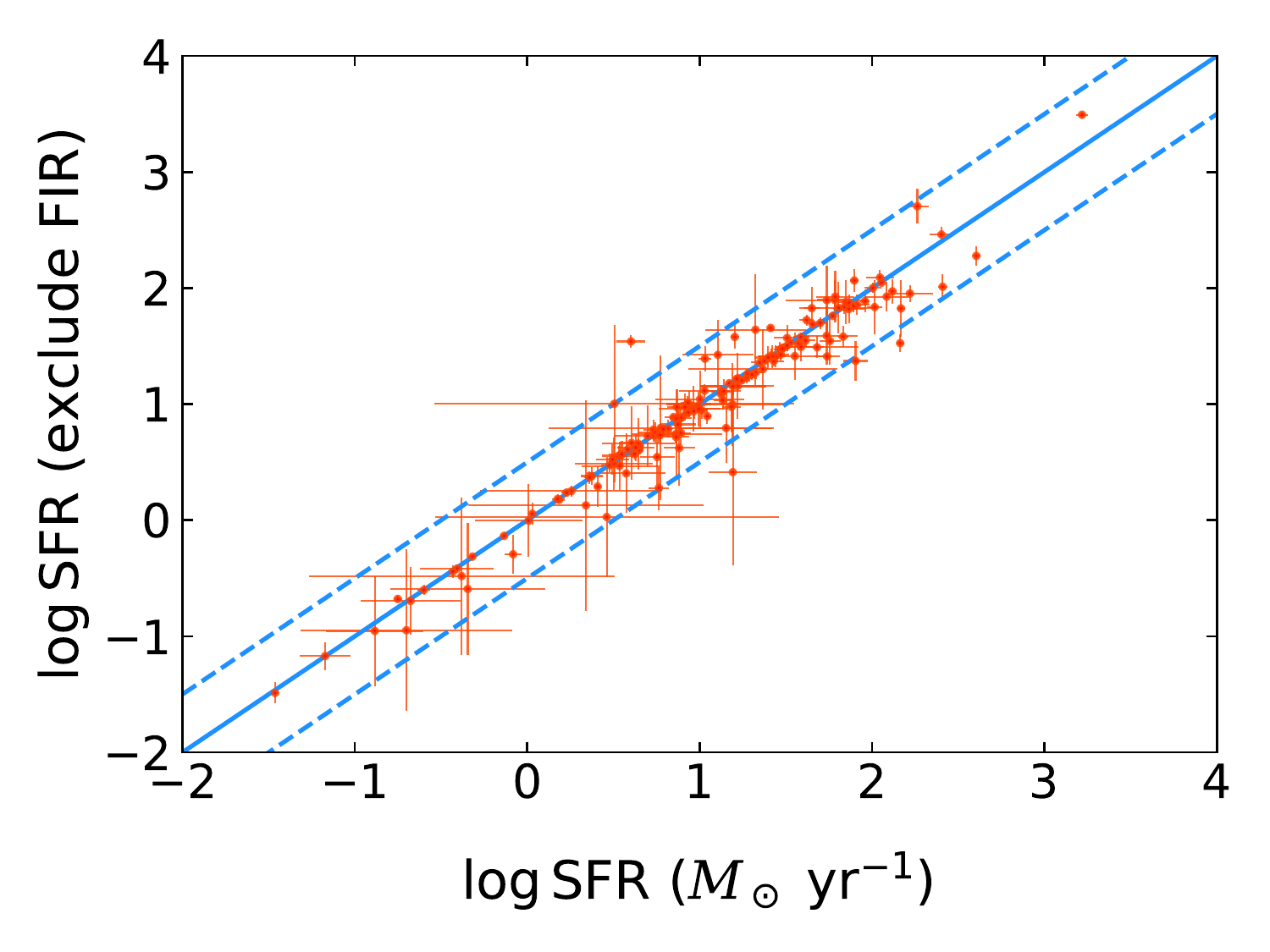}
\caption{Comparison of the SFRs obtained through including ($x$-axis) and excluding ($y$-axis) \herschel~fluxes in the SED fitting for \herschel-detected sources. The dashed blue lines mark the $\pm0.5$ dex regions that  deviate from the one-to-one correlation.}
\label{fig:cmp}
\end{figure} 

\section{SED-Fitting Method and Results}\label{sec:method}
To derive the host-galaxy properties for our sample, we perform multiwavelength SED fitting using X-CIGALE \citep{Yang2020} - a new release of the SED fitting code CIGALE \citep{Boquien2019}.
X-CIGALE has a few important improvements compared with CIGALE. First, it incorporates a new X-ray module which allows us to take advantage of the unique information of AGN intrinsic power provided by X-ray data, and fit SEDs from X-ray to infrared wavelengths. Second, it implements SKIRTOR \citep{Stalevski2012}, a two-phase clumpy torus model where the torus is illuminated by an anisotropic disk to account for AGN emission, which is more realistic than the previous smooth torus model \citep{Fritz2006} assumed in CIGALE and has been favored by recent simulations and observations \citep[e.g.,][]{Stalevski2012, Ichikawa2012, Xu2020}.

We adopt the delayed star formation history (SFH) which has a good performance of recovering the intrinsic galaxy parameters as verified via simulations \citep[e.g.,][]{Ciesla2015}. The BC03 stellar population models \citep{BC03} are adopted to produce galaxy SEDs by assuming the Chabrier initial mass function (IMF, \citealt{Chabrier2003}), which are then attenuated by the \cite{Calzetti2000} attenuation law, and re-radiated in IR using the \cite{Dale2014} dust templates. The modified SKIRTOR model \citep{Stalevski2012, Duras2017, Yang2020}, which consists of the direct disk radiation in the form of power laws and the re-radiation of the clumpy torus surrounding the central source, is adopted to model AGN emission from UV to IR wavelengths. Specifically, the disk SED is modeled as 
\begin{equation}
\lambda L_\lambda=\left\{
\begin{array}{lcl}
\lambda^2 & & {8 \leq \lambda < 50\ [\rm nm]}\\
\lambda^{0.8} & & {50 \leq \lambda < 125\ [\rm nm]}\\
\lambda^{-0.5} & & {125 \leq \lambda < 10^4\ [\rm nm]}\\
\lambda^{-5} & & {\lambda \geq 10^4\ [\rm nm]}
\end{array} \right.
\end{equation}
The attenuation of the torus is treated separately from that of the galaxy component. The output torus radiation is calculated based on the 3D radiative transfer code SKIRT \citep{Baes2011} which is dependent on the assumed geometric structure and density profile of the clumpy materials as well as the inclination angle.
The X-ray SED is modeled as a cutoff power law with the photon index being fixed to 1.8 during the fitting, which is the mean value for our sample derived through X-ray spectral fitting in paper I.  The X-ray emission is connected to other wavelengths via the $\alpha_{\rm ox} - L_{\rm 2500 \AA}$ relation expressed as $\alpha_{\rm ox} = -0.137 {\rm log}\,L_{\rm 2500 \AA} + 0.2638$ \citep[][]{Just2007}. Following \cite{Yang2020}, we adopt $|\Delta \alpha_{\rm ox}|$ which represents the deviation to the observed $\alpha_{\rm ox} - L_{\rm 2500 \AA}$ relation to be 0.2, corresponding to $\approx 2\sigma$ scatter of the $\alpha_{\rm ox} - L_{\rm 2500 \AA}$ relation. The summary of the main parameter ranges adopted in the fitting is presented in Table \ref{table:sed_para}. 

An example of our SED fitting results is displayed in Figure \ref{fig:sed_example}.  The AGN MIR luminosity is represented by the rest-frame 6 \um~luminosity derived from the decomposed AGN component. 
The galaxy stellar mass (i.e.,  {\tt bayes.stellar\_mass}) and SFR (i.e., {\tt bayes.sfh.sfr}) are adopted from CIGALE outputs. 

Note that the degeneracy between AGN and stellar components encountered during SED decomposition is potentially relevant when the FIR data are absent, and the constraints on SFR become poorer in this situation \cite[e.g.,][]{Gao2019}. To validate the usage of sources without solid \herschel~detections in our analyses, we carefully test the impact of the lack of FIR data as follows.
For \herschel-detected sources, we remove  all their \herschel~data points and re-fit the SEDs. We then compare the best-fit SFRs obtained from the with-\herschel~fitting to that from the without-\herschel~fitting as shown in Figure \ref{fig:cmp}. It can be seen that, the SFRs estimated from FIR-data-excluded SEDs are in good agreement with that derived from the whole SEDs. This result shows that benefiting from utilizing X-ray data in the fitting which provides a unique insight into the intrinsic AGN power, our good-photometric-coverage optical-to-MIR data are able to provide good constraints on SFR estimates even without resorting to FIR data. 

\begin{figure}
\centering
\includegraphics[width=\linewidth]{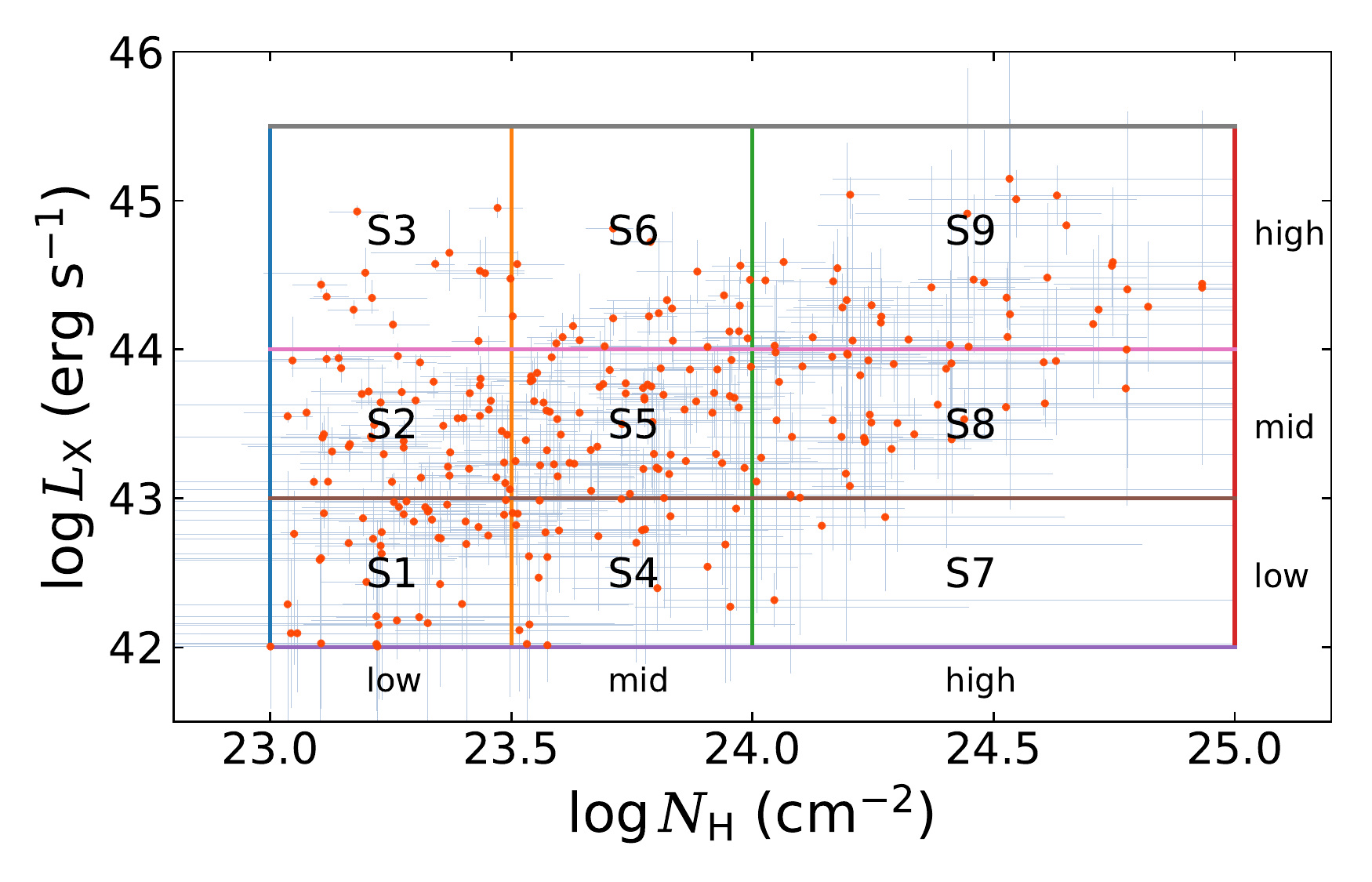}

\caption{Scatter plot of \loglx~vs. \lognh. The \lx-\nh~space is divided into nine bins (low/mid/high-\lx~vs. low/mid/high-\nh) which are annotated and used to calculate the median SEDs in Figure \ref{fig:sed}.}
\label{fig:divide_bin}
\end{figure} 

\begin{figure*}
\centering
\includegraphics[width=\linewidth]{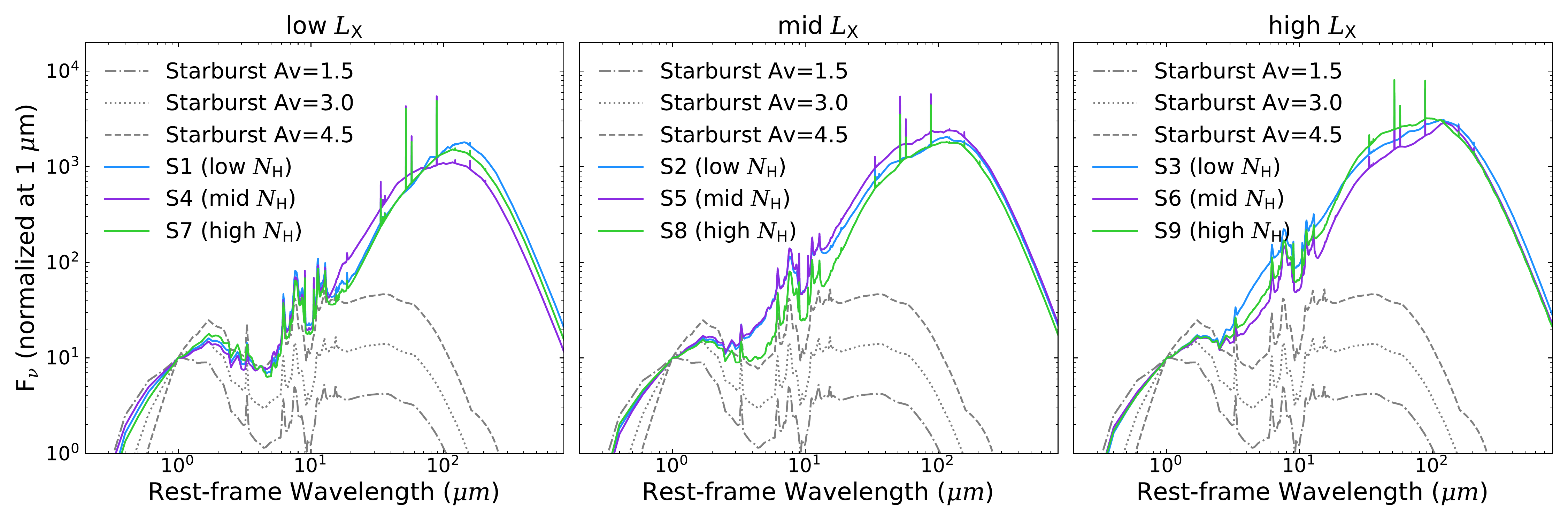}
\includegraphics[width=\linewidth]{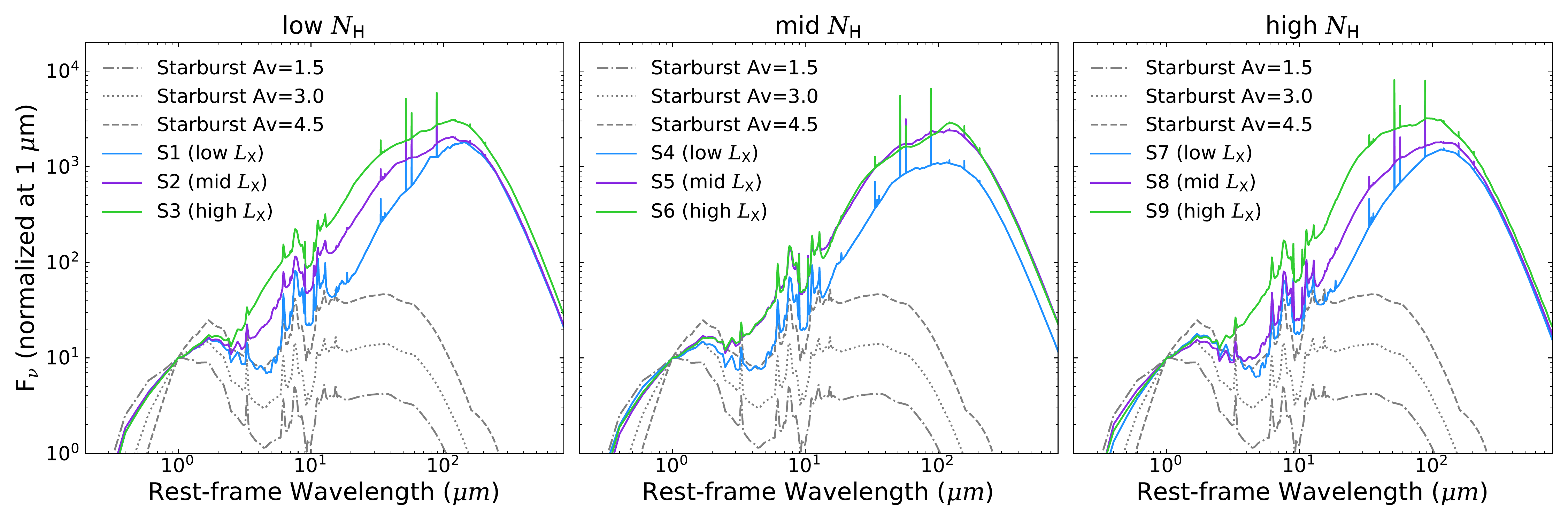}

\caption{Composite (stellar + AGN) SEDs for highly obscured AGNs in different \lx~and \nh~bins (i.e., S1 -- S9) defined in Figure \ref{fig:divide_bin}. Top (Bottom): Comparison of SED shapes at a given \lx~(\nh) bin.
Each gray curve shows the M82 galaxy template that is attenuated using a specified extinction value (without adding the dust re-radiation component in FIR).}
\label{fig:sed}
\end{figure*}

\section{Revisiting the Optical/IR-Selection Methods}\label{sec:IR}
We first explore the dependences of SED shapes on AGN physical properties, specifically, the X-ray luminosity and obscuring column density. We divide our highly obscured sample into nine \lx~and \nh~bins (see Figure \ref{fig:divide_bin}) and calculate the  median composite (AGN + stellar) SED in each bin. The individual SEDs are normalized at rest-frame 1~\um~before calculating the median SED and the results are displayed in Figure \ref{fig:sed}. 

Comparing the results in different \nh~and \lx~bins, we find that the dependence of the composite  SED shape on \nh~is not as sensitive as that on \lx~at optical-to-MIR wavelengths, which is traced by the large differences between low-luminosity and luminous sources in a given \nh~bin (e.g., S1 vs. S3). We also show the M82 starburst galaxy template being attenuated by three different extinction values in Figure \ref{fig:sed} for comparison (the re-radiation at FIR is not included). The composite SED shapes for luminous sources are similar to those for typical IR-bright power-law AGNs \citep[e.g.,][]{Donley2012}; but for low-luminosity objects, the prominent NIR bump makes their NIR-to-MIR SEDs more similar to those of galaxies whose emission is dominated by dust-obscured star formation \citep[e.g.,][]{Riguccini2015}. 
This overall similarity makes it challenging to identify low-luminosity, highly obscured AGNs using pure SED diagnostics. 

Several works have been devoted to using optical and IR colors to select obscured AGN candidates, such as IR-excess methods \citep[e.g.,][]{Daddi2007, Alexander2008, Luo2011}, \textit{WISE}-color selection methods \citep[e.g.,][]{Tsai2015, Fan2016a, Glikman2018} and IR-to-optical flux-ratio diagnostics \citep[e.g.,][]{Fiore2008}; and the studies based on these selection criteria have yielded remarkable insights into our understanding of the obscured AGN population. However, the optical/IR SED-based methods may be biased against low-luminosity AGNs. In the following sections, we will discuss several selection methods in detail. We do not intend to directly quantify the completeness and reliability of each method, but mainly focus on what implications we can deduce by comparing the properties of sources selected using different diagnostics in order to better understand the highly obscured AGN population. 

\subsection{Can IRAC Colors Effectively Identify Luminous Highly Obscured AGNs?}\label{subsec:irac}

\begin{figure*}
\centering
\includegraphics[width=1.0\linewidth]{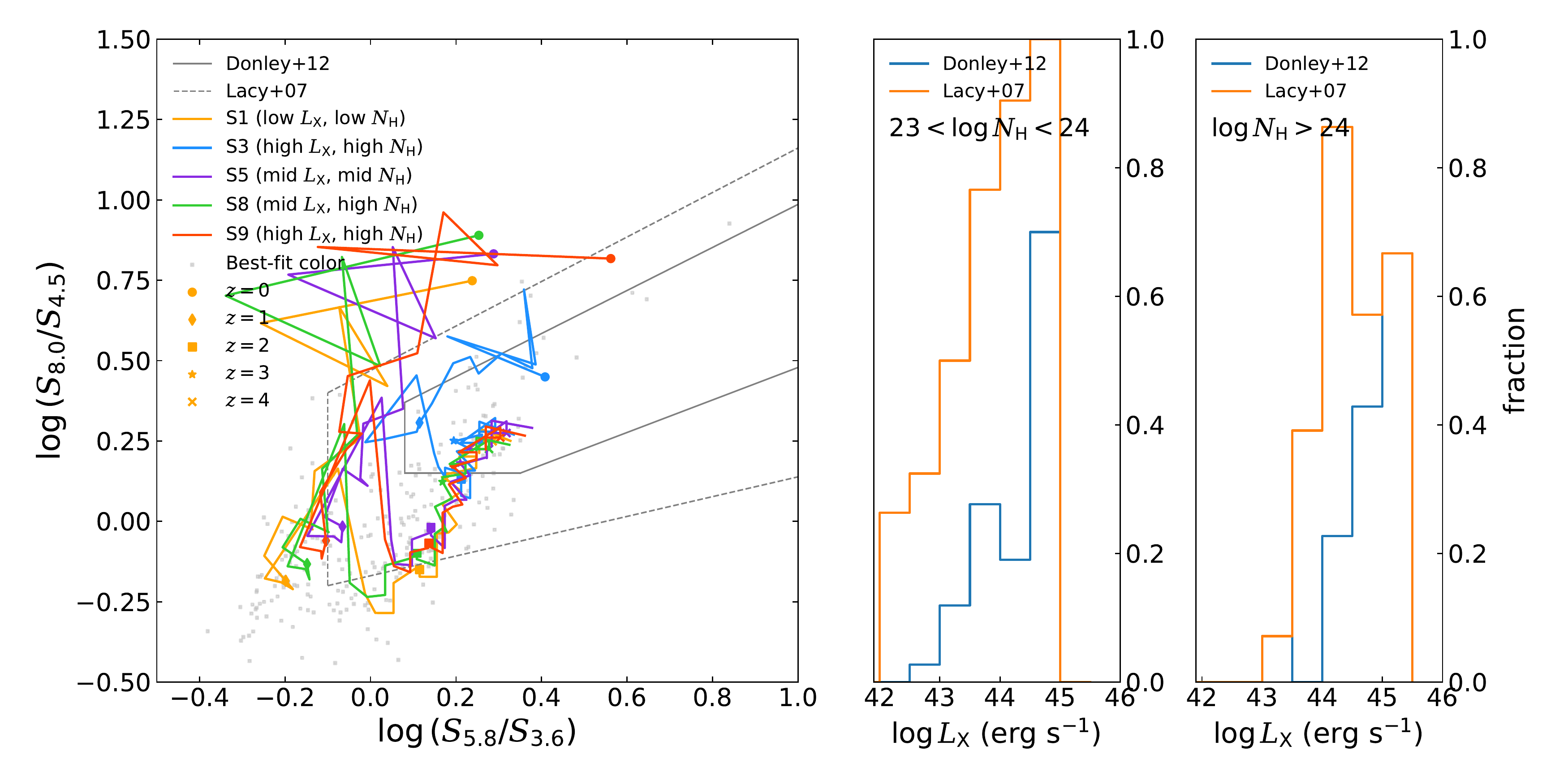}
\caption{Left: IRAC color-color diagram for 247 highly obscured AGNs that are detected in all the four IRAC bands. The gray points represent each individual source. The solid segmented lines represent the color evolutionary tracks (in steps of $\Delta z=0.1$) in five \lx-\nh~bins (i.e., S1, S3, S5, S8 and S9; see Figure \ref{fig:divide_bin}) calculated from the median composite SEDs in Figure \ref{fig:sed}. We denote the five redshift nodes from $z=0$--4 using colored markers. Right: Fractions of our highly obscured AGNs being identified as AGNs by the \cite{Donley2012} and \cite{Lacy2007} criteria (shown as the wedges in the left panel) as a function of \lx~for the $\cm < \nh < \ct$ and $\nh > \ct$ bins, respectively. }
\label{fig:irac}
\end{figure*} 

Among the IR-AGN selection methods, IRAC color is a powerful tool to select large samples of  luminous AGN candidates (\citealt{Stern2005}, \citealt{Lacy2007}, hereafter L07; \citealt{Donley2012}, hereafter D12). The most promising aspect of this method compared to X-ray selections \citep[e.g.,][]{Xue2011,Xue2016,Luo2017,Xue2017} is its ability to recover the most heavily obscured sources that are often not detected in X-rays (e.g., 62\% of IRAC-selected AGNs do not have X-ray counterparts in D12, which is attributed to heavy X-ray obscuration).

In Figure \ref{fig:irac} we plot 247 highly obscured AGNs which are detected/covered in all the four IRAC bands on the IRAC color-color diagram, as well as the color evolutionary tracks at different redshifts calculated from the median composite SEDs in Figure \ref{fig:sed}.
The color evolutionary tracks at $z>1$ are located well within the L07 wedge, suggesting that the L07 criterion should be able to identify highly obscured AGNs efficiently. In contrast, almost all the color tracks at $z<3$ avoid the refined D12 wedge, suggesting that the X-ray selected highly obscured AGNs will generally be missed by the D12 selection criterion.

When showing in the right panel of Figure \ref{fig:irac} the fractions of our sources being identified as AGNs by the L07 and D12 criteria as a function of \lx~in two \nh~bins (corresponding to highly obscured CN and CT sources, respectively), we find that, at $\loglx > 44.5\ \ergs$, the L07 and D12 criteria can recover a substantial fraction of luminous, X-ray-selected highly obscured AGNs with $\lognhbar \sim 23.5\ \nhu$. Such a value is in good agreement with the average column density ($\lognh \sim 23.5\pm0.4\ \nhu$) derived through stacking X-ray-undetected IRAC-selected AGNs in D12 using shallower X-ray data. 

However, at $\loglx < 44.5\ \ergs$, the selected source fraction using the D12 criterion dramatically drops even within the range of $44\ \ergs < \loglx < 44.5\ \ergs$, which suggests that it is still incomplete in selecting highly obscured AGNs even for the luminous population \citep[e.g.,][]{Kirkpatrick2017}. The \logmbar~and \logsfrbar~values for the D12-missed luminous highly obscured AGNs are 10.9~\msun~and 1.3~\msun/yr, which are lower than the D12-selected sources with  $\logmbar = 11.2\ \msun$ and $\logsfrbar = 1.7\ \msun$/yr. Therefore, we suppose that the host-galaxy contamination should not be the main reason responsible for missing a large population of luminous highly obscured AGNs. The lower average redshift for the missed sources ($\zbar \sim 2.3$) than that of the selected sources ($\zbar \sim 3.3$) may partly explain the reduced selection efficiency, as can be seen from the color evolutionary tracks. However, we notice that the average redshift for the missed sources is similar to that of IRAC-selected AGNs in D12 ($z\sim1.8$ and $z\sim2.1$ for X-ray detected and non-detected sources, respectively).
Therefore, the reason that these X-ray luminous highly obscured AGNs are missed is likely that they are intrinsically fainter in MIR. This can be due to the lower dust contents and/or CFs of the tori  makes their SEDs more similar to that of star-forming galaxies, as verified by their lower \loglmirbar~value ($\sim$44.1~\ergs) compared to that of the selected sources ($\sim$45.1~\ergs) while the average X-ray luminosities for the two populations are similar ($\loglxbar \sim 44.4\ \ergs$). 

For low-luminosity bins, the D12 criterion misses the majority of our sources since the MIR SEDs of low-luminosity AGNs are largely contaminated by the host-galaxy emission (see Figure \ref{fig:sed}); the more-relaxed L07 criterion maintains a relatively high completeness, but suffers large contamination from distant star-forming and starburst galaxies \citep[e.g.,][]{Donley2012}.

\subsection{Is the High Ratio of \fmirr~an Efficient Method to Select Highly Obscured AGNs?}\label{subsec:f24}

\begin{figure*}
\centering
\includegraphics[width=0.49\linewidth]{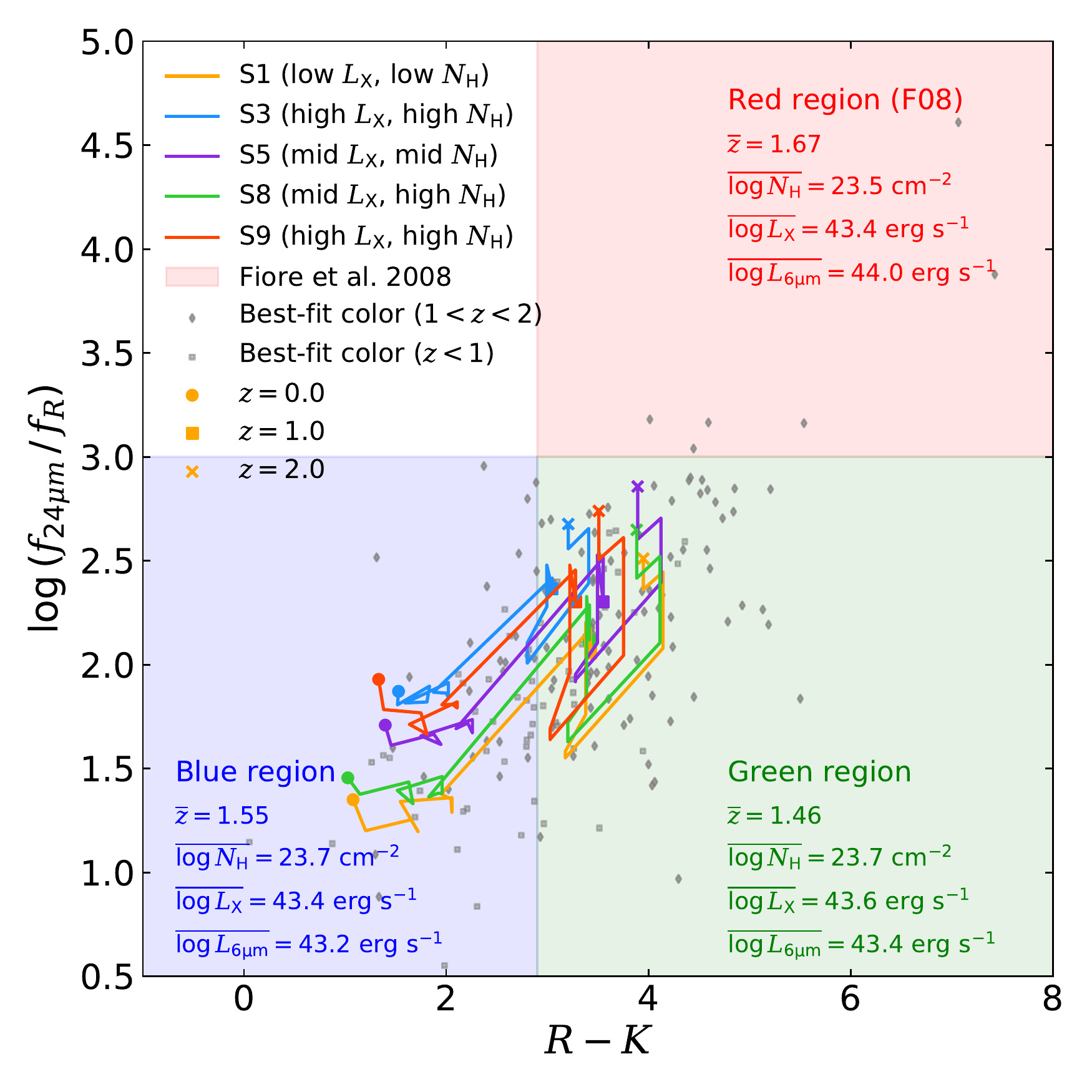}
\includegraphics[width=0.49\linewidth]{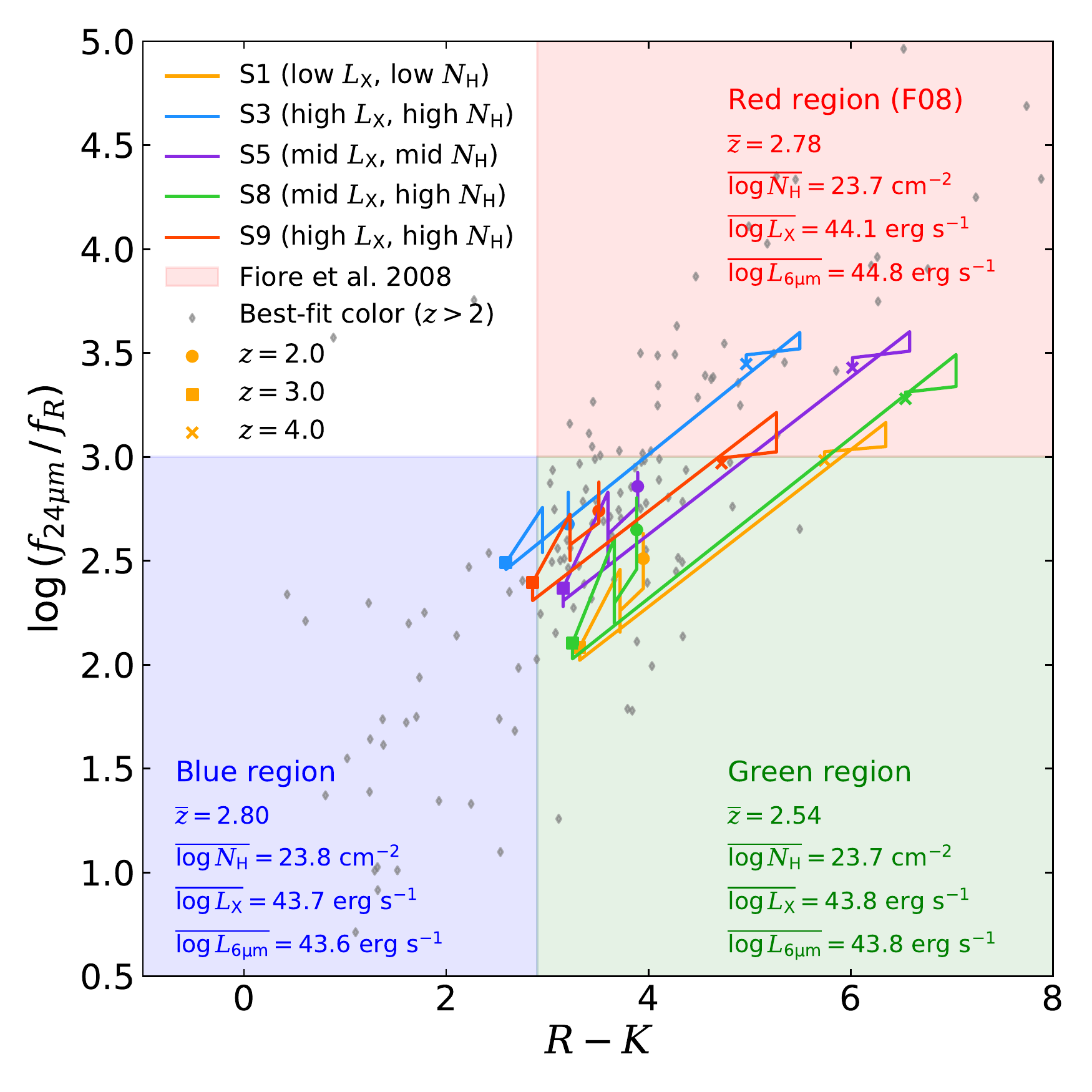}
\caption{\fmirr~vs. $R-K$ (in AB magnitudes) color-color diagrams for highly obscured AGNs at $z<2$ (Left) and $z>2$ (Right), respectively. The individual points represent the best-fit colors for our sources calculated from the model-predicted fluxes (see the red circles in Figure \ref{fig:sed_example}). 
The \zbar, \lognhbar, \loglxbar, \loglmirbar~values for each region are listed for comparison. Note that for the $z < 2$ diagram, since there are no sources at $z < 1$ located in the red region, the average values are calculated for sources with $1 < z < 2$.
The solid segmented lines represent the color evolutionary tracks in five \lx-\nh~bins (see Figure \ref{fig:divide_bin}) calculated from the median composite SEDs in Figure \ref{fig:sed}. In each panel, we label three redshift nodes for the color evolution tracks. }
\label{fig:rk24}
\end{figure*} 

Because of large obscuration in highly obscured AGNs, the bulk of UV-optical photons are absorbed and re-emitted in the IR with a peak at MIR. In addition, obscured AGNs tend to have red colors \citep[e.g,][]{Brusa2005}. Consequently, a large MIR (e.g., \spitzer~24~\um) to optical (e.g., $R$-band) flux ratio combined with a red color (e.g., $R-K$) is expected to be a good tracer of high-level obscuration.
 
\cite{Fiore2008} (hereafter F08) applied the $\fmirr >1000$ and $R-K > 4.5$ (in Vega magnitudes, corresponding to 2.86 in AB magnitudes) criteria to the GOODS-MUSIC catalog \citep{Grazian2006} to select the ``missing'' highly obscured AGN candidates at $z\sim 1.5-2.5$ that complement X-ray selections. For the 22 X-ray-detected sources in the 1~Ms CDF-S \citep{Giacconi2002}, the hardness-ratio analysis indicates that they are obscured AGNs with $\nh > 10^{22}\ \nhu$. For the 111 X-ray-undetected sources, the combined stacking analysis and Monte Carlo simulation show that  $\sim 80\%$ of them are possibly highly obscured AGNs. In the era of the 7~Ms CDF-S, with the additional 6~Ms exposure which significantly improves the detectability of heavily obscured sources that are hidden in the previous 1~Ms CDF-S data, we are able to investigate this method in more detail.

In Figure \ref{fig:rk24} we plot our highly obscured AGNs on the \fmirr~versus $R-K$ (in AB magnitudes) digram using the fluxes predicted at filter wavelengths (i.e., red filled circles in Figure \ref{fig:sed_example}), as well as the color evolutionary tracks similar to those in Figure \ref{fig:irac}. The choice of using model-predicted fluxes instead of observed fluxes here is to enlarge the sample being investigated (i.e., the whole sample can be plotted and we can include each sources, even the 101 sources not covered in all bands, in the red, green or blue populations defined by the shaded regions in Figure \ref{fig:rk24}). We note that using actual observed fluxes to derive colors does not affect our conclusion qualitatively, although the exact values of colors will be slightly different.

The expected correlation between \fmirr~and $R-K$ color can be clearly seen \citep[e.g.,][]{Fiore2008}, and our sources indeed have a much redder mean color (i.e., $\Delta \overline{R-K} = 1.9$) compared to the remaining sources in the SIMPLE survey \citep{Damen2011}.

There are 46 (16\%) sources located in the red region defined by F08 (i.e., $\fmirr > 1000$ and $R-K > 2.86$), with 40 of them at $z>2$ and the remaining at $1<z<2$, indicating that this criterion can indeed select heavily obscured AGNs. However, the average redshift for our ``red'' sources ($\zbar=2.7$) is significantly higher than that of the highly obscured AGN candidates selected in F08 which peaks at $z = 1.5-2.0$. The very low fraction of our sources residing in the red region and the large redshift discrepancy suggest either large incompleteness of this method \citep{Comastri2011, Brightman2012} or an essential difference between X-ray- and IR-selected populations \citep[e.g.,][]{Hickox2009}. 

Note that the \logsfrbar~of red sources is slightly lower (i.e., $\Delta \logsfrbar \sim -0.2$ dex) than that of blue ones (i.e., having $\fmirr < 1000$ and $R-K < 2.86$) with matched redshifts, hence the increased $f_{24 \um}$ of red sources is not primarily caused by the enhanced star formation, but should be related to the central AGN. Even if we only consider the most-luminous sources (i.e., with $\loglx > 44\ \ergs$) to avoid host contamination to the observed colors, most of them (53/76) still avoid the red region.

To understand the differences between red, blue and green (i.e., having $\fmirr < 1000$ and $R-K > 2.86$) sources, we annotate their average source properties in Figure \ref{fig:rk24}. It can be seen that, at similar redshifts, the average \nh~and \lx~for the three source populations are roughly the same, but red sources have significantly higher \lmirbar~than those of blue and green sources.
Aside from the diverse galaxy contributions to the observed colors, another explanation for the widely distributed colors of our sources in a given redshift bin could be that the dust contents and CFs of the tori for blue and green sources are smaller than that of red ones, resulting in weaker reprocessed MIR emission and smaller \fmirr. 
Alternatively, if a significant portion of the heavy X-ray obscuration is contributed by dust-free materials such as broad-line region (BLR) gas and/or disk wind \citep[e.g., ][]{Burtscher2016, Liu2018, Ichikawa2019}, the UV-optical continuum will not be significantly attenuated, leading to smaller values of \fmirr~and $R-K$. It is also possible that the interstellar medium (ISM) may contribute significantly to X-ray absorption even up to $\nh > \cm$ for high-redshift gas-rich galaxies \citep[e.g.,][]{Gilli2014, Shu2018, Circosta2019, DAmato2020}; if so, since the dust temperature in the ISM is much lower than that in the torus, the reprocessed emission will peak at longer wavelengths (e.g., FIR-to-submm), thus the \fmirr~value may not be that large. Indeed, the $z=4.75$ CT AGN \citep[][]{Gilli2011} reported in the 4~Ms CDF-S \citep[XID 403,][]{Xue2011}, whose ISM in the central starburst region ($r_{\rm half} \sim 0.9\pm0.3\ \rm kpc$) is able to produce $\nh \sim (0.3-1.1) \times \ct$ as revealed by ALMA observations \citep{Gilli2014}, does have a very low value of \fmirr~=168, especially considering that its very high redshift is supposed to make it easier to fulfill the F08 criteria (see the color evolutionary tracks in the right panel of Figure \ref{fig:rk24}).

In conclusion, we find that the heaviest X-ray obscuration is not equivalent to extremely large \fmirr~and the reddest color, possibly owing to the diverse properties of obscuring materials (e.g., different CFs, gas/dust contents, and \nh~distributions), complex origins of the X-ray obscuration along our sightline (e.g., X-rays absorbed by dust-free BLR gas, disk wind, dusty torus and/or ISM) as well as galaxy contamination to the observed colors.

\begin{figure}
\centering
\includegraphics[width=\linewidth]{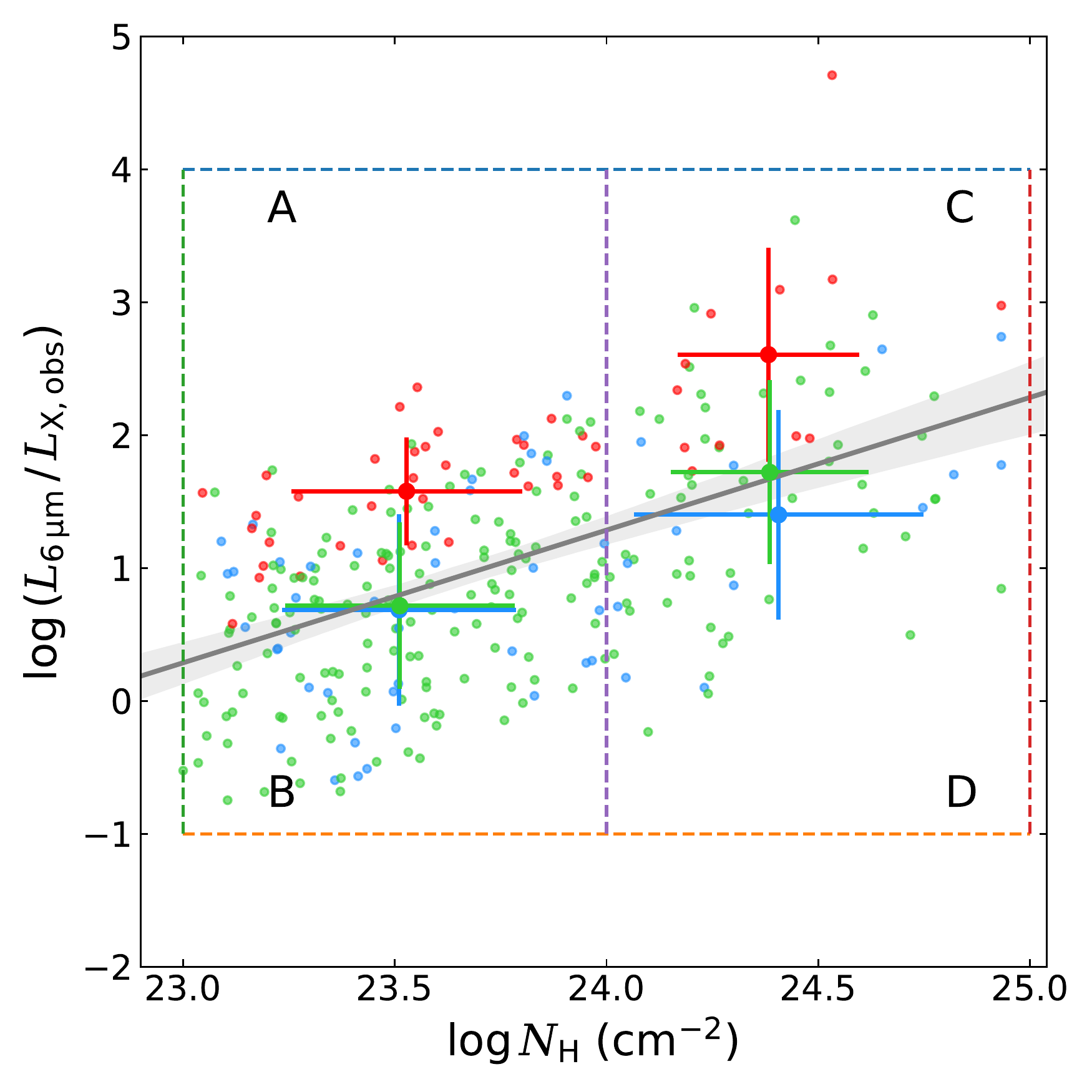}
\includegraphics[width=\linewidth]{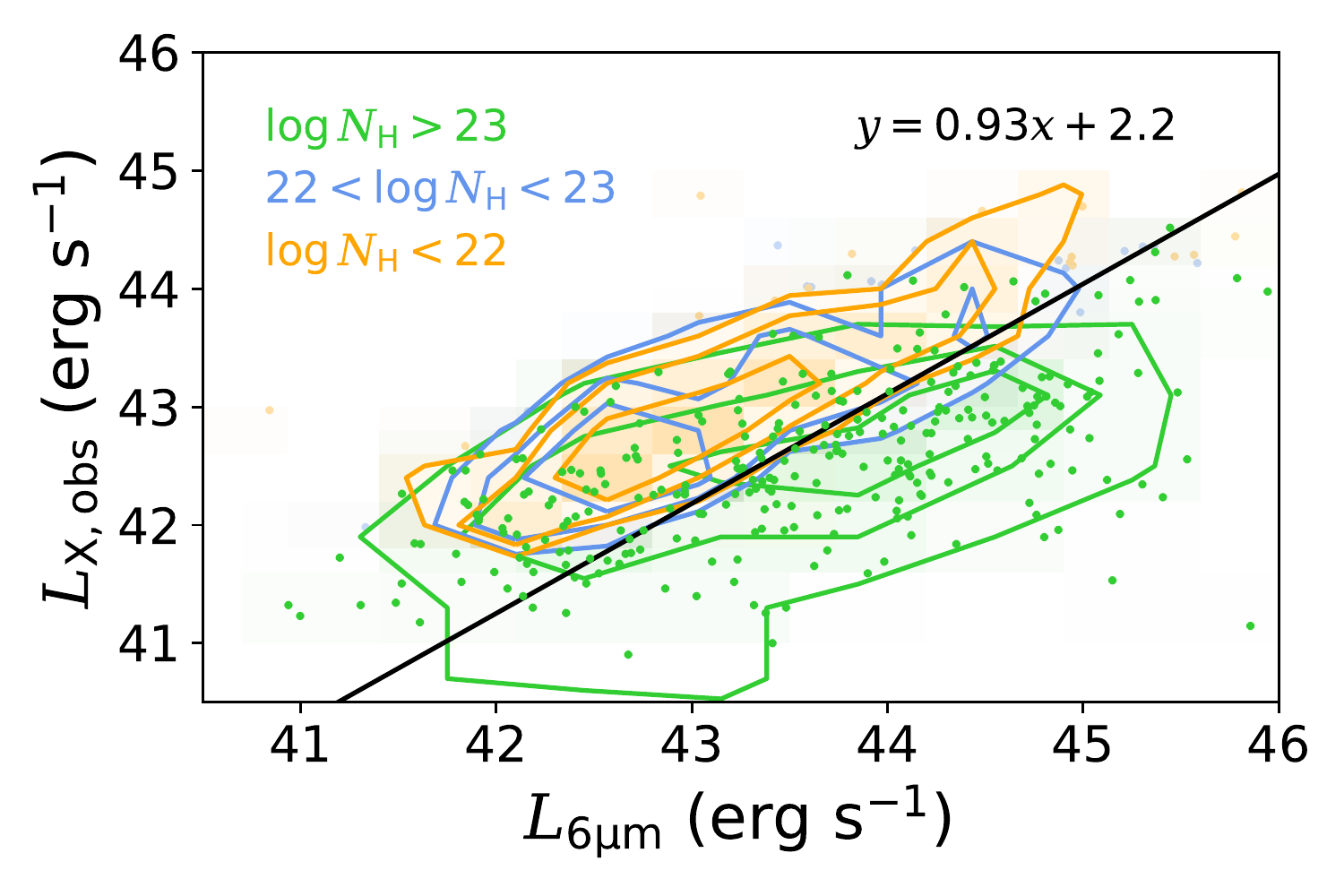}

\caption{Top: Dependence of \lratio~on \nh~for highly obscured AGNs. 
Symbol colors indicate the source locations in the red, blue and green regions that are defined in Figure \ref{fig:rk24}, respectively; accordingly, 
large crosses represent the mean \lratio~and \nh~in the CT and highly obscured CN regimes with corresponding error bars being the standard dispersions of \lratio~and \nh, respectively.
The gray line and associated shaded region are the best-fit correlation between \lratio~and \nh~and the $1\sigma$ uncertainty, respectively; this best-fit line and the vertical \nh=\ct~line divide this panel into four regions (i.e., A--D). 
Highly obscured AGNs with small CFs and hard spectral shapes may have small values of \lratio~(i.e., regions B and D), and high-CF less-obscured AGNs with soft X-ray spectra may have \lratio~as large as CT AGNs (i.e., region A vs. region D). 
Bottom: \lxobs~vs. \lmir~relation in three \nh~ranges. The black line represents the optimal boundary for separating highly obscured and less obscured AGNs derived through the linear Support Vector Machine algorithm.
}
\label{fig:est_nh}
\end{figure} 

\subsection{Can the Value of \lratio~be Used as a Reliable Indicator of \nh?}
\label{subsec:est_nh}

Since the AGN MIR emission produced by the absorption and re-radiation of UV-to-optical photons is largely unaffected by dust attenuation, whereas X-ray photons will be significantly absorbed when \nh~reaches the highly obscured regime, a large ratio of the MIR luminosity to observed X-ray luminosity (\lratio) has been widely adopted as an indicator of heavy obscuration \citep[e.g.,][]{Alexander2008, DelMoro2013, Rovilos2014, DelMoro2016, Corral2016}.

In Figure \ref{fig:est_nh} we show the dependence of \lratio~on \nh~for our sample. We confirm that there is a positive correlation between the two parameters (with Spearman's $\rho = 0.40$ and $p \ll 0.001$), albeit with large dispersion. Considering the theoretical argument proposed by \cite{Yaqoob2011} that \lratio~is more sensitive to the torus CF and the incident X-ray continuum shape, rather than \nh, it is possible that highly obscured AGNs with small CFs and hard spectral shapes may have lower \lratio~values (see the sources in regions B and D in Figure \ref{fig:est_nh} that lie below the best-fit line); and high-CF less-obscured AGNs with soft X-ray spectra may have \lratio~values as large as CT AGNs (see the sources in region A vs. those in region D). 

These statements are supported by the result that when we plot in Figure \ref{fig:est_nh} the \loglratiobar~and \lognhbar~values for the $z>1$ red, green and blue populations defined in Figure~\ref{fig:rk24} and Section \ref{subsec:f24}, it can be clearly seen that red sources show the highest \lratiobar~at a given \nh, consistent with a scenario that they are deeply buried by plentiful dusty materials. While for blue and green sources, the dust contents and CFs might be lower, resulting in smaller values of \lratio~even though the levels of their line-of-sight (LOS) X-ray obscuration are indistinguishable from red sources. Therefore, we conclude that a simple \lratio~value alone is not sufficient to identify CT AGNs (also see \citealt{Georgantopoulos2011}). 

Even so, the \lratio~value may still be useful to distinguish highly obscured and less obscured AGNs. In the bottom panel of Figure \ref{fig:est_nh} we show the \lxobs~vs. \lmir~relation at three \nh~ranges by including less obscured AGNs in the plot. The majority ($\sim60\%$) of highly obscured AGNs are separated from less obscured ones, while sources with $\lognh < 22\ \nhu$ are mixed with those with $22\ \nhu < \lognh < 23\ \nhu$ due to the fact that the rest-frame 2-10 keV flux is not significantly absorbed in the Compton-thin regime. We use the linear Support Vector Machine algorithm built in the Python package {\tt scikit-learn} to derive the boundary line between highly obscured and less obscured AGNs. The optimal boundary is shown in black line, parameterized as 
\begin{equation}
\lxobs = 0.93 \times \lmir + 2.2.
\end{equation}
This boundary line can be served as a complementary method to the hardness ratio criterion we presented in Section 4.3 of paper I to select highly obscured AGNs.

\section{Are Highly Obscured AGNs the Missing Link in the Merger Models?}
\label{sec:missing_link}

\subsection{Do Highly Obscured AGNs Show Enhanced Star Formation Activities?}
\label{subsec:msq}

\begin{figure*}
\centering
\includegraphics[width=1.0\linewidth]{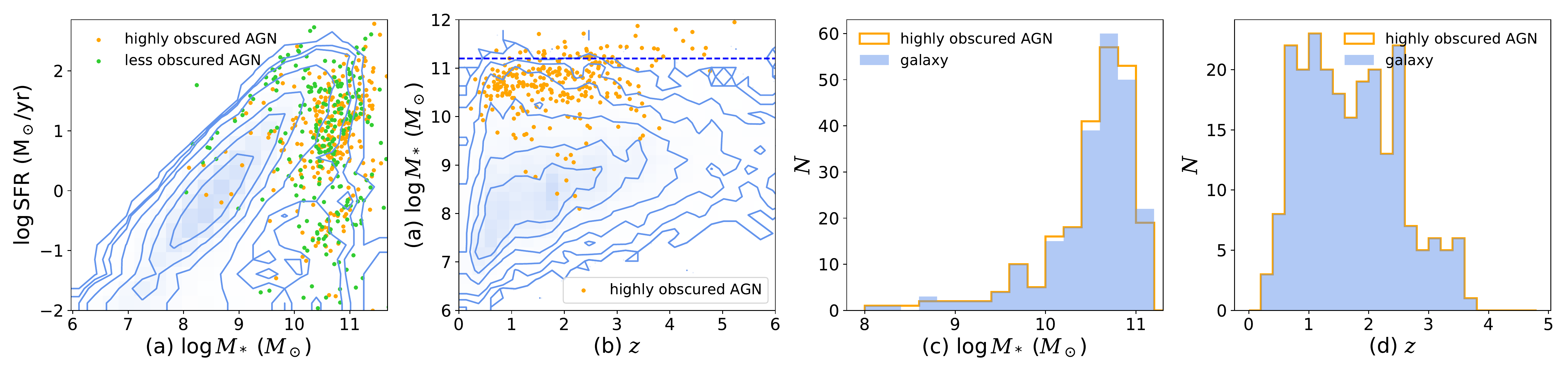}
\caption{ (a) SFR vs. $M_*$ relation for our sample of highly obscured and less obscured AGNs compared to that of non-active galaxies (blue contours) in \cite{Santini2015}.  (b) $M_*$ vs. $z$ for highly obscured AGNs relative to non-active galaxies. The horizontal dashed line marks out stellar mass cut at log~$M_*/M_\odot$=11.2. (c, d) $M_*$ and $z$ distributions for typical randomly-selected highly obscured AGNs and their corresponding control galaxies from our sampling procedure.}
\label{fig:msq}
\end{figure*}

To evaluate the star-forming activity of our highly obscured AGN sample in the context of the general galaxy population, we construct a control non-active galaxy sample from \cite{Santini2015} (hereafter S15) in which the SED-fitting results for 34,929 galaxies from ten independent teams adopting different model configurations are available. The X-ray AGNs identified in the 7~Ms~CDF-S catalog are excluded. Following \cite{Yang2017}, we adopt the median values of \ms~and SFR reported from teams $2a_\tau$ , $6a_\tau$, $11a_\tau$, $13a_\tau$ and $14a$ in the following analyses, all of which assumed the same BC03 stellar templates and Chabrier IMF as in this paper. 

Note that although S15 does not consider the AGN component in the SED fitting,  their results may still provide reliable mass estimates for highly obscured AGNs with moderate luminosities, as their rest-frame optical-to-NIR SEDs (which are the most important to constrain \ms) are largely dominated by the galaxy component \citep[e.g.,][]{Luo2010, Xue2010}. Therefore, we compare our \ms~estimates with S15 for the 82 common sources in the two works with redshift difference $\Delta z < 0.05$. The derived $\overline{\Delta \logms}$ between the two works is $0.06 \pm 0.02$ dex, suggesting that there is no significant systematic bias induced by the different SED-fitting approaches.

In Figure \ref{fig:msq}a we plot our highly obscured AGNs in the SFR vs. $M_*$ plane. Also shown are less obscured AGNs with a roughly matched redshift distribution (see Figure \ref{fig:sample}) and normal galaxies from S15 at similar redshifts. The distributions of SFR of the two AGN populations suggest that they are mainly hosted by star-forming galaxies, and there is no noticeable enhancement of star-forming activity in highly obscured AGNs than less obscured AGNs \citep[e.g.,][]{Zou2019, Suh2019}. 

To further control the redshift and $M_*$ dependence of SFR, we divide the $M_* - z$ space (Figure \ref{fig:msq}b) into a series of subgrids with $\Delta M_* = 0.2$ dex and $\Delta z = 0.2$. In each subgrid, the number of highly obscured AGNs is denoted as $N_i$ and we randomly select $N_i$ highly obscured AGNs and $N_i$ normal galaxies allowing duplication. By repeating the procedure in each subgrid, new highly obscured AGN and normal galaxy samples with matched $z$ and $M_*$ distributions can be constructed. Note that the fraction of galaxies hosting an AGN dramatically increases with \ms~\citep[e.g.,][]{Xue2010, Yang2017, Yang2018}, thus at the highest mass end we may not be able to find a sufficient number of control galaxies that do not contain an active SMBH, as is the case for our $\logms > 11.2$ \msun~sources (see Figure \ref{fig:msq}b). Therefore, we restrict our sampling pocedure to $\logms < 11.2\ \msun$ which accounts for $85\%$ of our sample. 
We perform the above procedures 1000 times and show one example of the distributions of $M_*$ and $z$ for a randomly-selected AGN sample and a control galaxy sample in Figures \ref{fig:msq}c and \ref{fig:msq}d. As can be seen, the $M_*$ and $z$ distributions have been well controlled. 
The random samples vary every time 
we repeat the sampling procedure. For each sampling, we calculate the 20th, 50th and 80th percentiles from the SFR distributions of the randomly selected AGNs and control galaxies. The average SFR (\logsfrbar) at each percentile is calculated by averaging the values of the 1000 random samples and the results are summarized in Table \ref{table:delta_sfr}. The respective $1\sigma$ uncertainty is calculated from the 84th-percentile and the 16th-percentile of the corresponding resampled parameter distribution. 

\begin{table*}
\centering
\caption{Comparison of star formation properties between highly obscured AGNs and their $M_*$- and $z$-controlled normal galaxies.}
\begin{tabular}{c c c c c c c}
\hline
\hline
Sample & $\overline{\logsfr_{\rm agn, 50th}}$ & $\overline{\logsfr_{\rm gal, 50th}}$ & $\overline{\logsfr_{\rm agn, 20th}}$ & $\overline{\logsfr_{\rm gal, 20th}}$ & $\overline{\logsfr_{\rm agn, 80th}}$ & $\overline{\logsfr_{\rm gal, 80th}}$ \\
\hline

Total & $0.87_{-0.02}^{+0.02}$ & $0.90_{-0.13}^{+0.11}$ & $0.20_{-0.15}^{+0.12}$ & $-0.63_{-0.14}^{+0.17}$ & $1.40_{-0.07}^{+0.04}$ & $1.70_{-0.05}^{+0.06}$\\

high \lx & $0.91_{-0.03}^{+0.02}$ & $0.92_{-0.18}^{+0.17}$ & $0.49_{-0.11}^{+0.06}$ & $-0.73_{-0.16}^{+0.20}$ & $1.31_{-0.07}^{+0.04}$ & $1.75_{-0.11}^{+0.06}$\\

low \lx & $0.75_{-0.08}^{+0.09}$ & $0.87_{-0.17}^{+0.17}$ & $0.00_{-0.14}^{+0.06}$ & $-0.46_{-0.30}^{+0.17}$ & $1.40_{-0.07}^{+0.04}$ & $1.65_{-0.06}^{+0.07}$\\
\hline

\hline
\end{tabular}
\label{table:delta_sfr}
\end{table*}

The \logsfrbarm~for highly obscured AGN hosts is found to be consistent with that of control galaxies ($\deltalogsfrbarm = -0.03_{-0.12}^{+0.12}$). Such a result also holds for both low- ($\deltalogsfrbarm = -0.11_{-0.19}^{+0.15}$) and high-luminosity ($\deltalogsfrbarm = -0.02_{-0.16}^{+0.20}$) AGNs if we split the sample into two subsamples based on the median \lx~at each redshift grid. The \logsfrbarh~for highly obscured AGN hosts appears to be lower than that of control galaxies with $\deltalogsfrbarh = -0.30_{-0.07}^{+0.08}$. These results suggest that the star-forming activity of highly obscured AGNs is not enhanced with respect to normal star-forming galaxies.

However, a deficiency of quiescent hosts among highly obscured AGNs can be clearly seen with $\deltalogsfrbarl \approx 0.83_{-0.20}^{+0.17}$. 
As a result, the SFR distribution of highly obscured AGNs appears to be less diverse and more main-sequence like than that of normal galaxies \citep[e.g.,][]{Bernhard2019}, suggesting that the sufficient cold gas supply that is responsible for sustaining star formation may also be an important factor in triggering highly obscured AGNs. On the other side, the indistinguishable \logsfrbarm~for highly obscured AGN hosts to that of control galaxies suggests that, unlike the significant populations of IR-selected ultraluminous IR galaxies and hot dust-obscured galaxies which are generally believed to hold both highly obscured AGN \citep[e.g.,][]{Vito2018} and enhanced starburst activity as a consequence of mergers \citep[e.g.,][]{Farrah2003, Fan2016a, Fan2016b}, there is no evidence supporting that the presence of X-ray-selected highly obscured AGNs is more frequently connected to violent, possibly merger-driven starburst activities \citep[e.g.,][]{Georgantopoulos2013, Lanzuisi2015, Suh2017}.

\subsection{Are AGN Activity and Obscuration Linked with Host-galaxy Properties in Highly Obscured AGNs?}
\label{subsec:link}

Many studies have explored the correlations between AGN and host-galaxy properties (e.g., \lx, \nh~vs. $M_*$, SFR) in a variety of redshift ranges, but there have been considerable debates about whether AGN activity and obscuration are linked with galaxy-wide star formation \citep[e.g.,][]{Lutz2010, Shao2010, Page2012, Harrison2012, Stanley2015, Lanzuisi2017, Dai2018, Schulze2019}, as well as whether $M_*$ or host-galaxy compactness is a more fundamental factor that governs the average black-hole accretion rate (BHAR) \citep[e.g.,][]{Yang2017, Yang2018, Fornasini2018, Ni2019}. In particular, \cite{Yang2019} presented an attractive scenario in which the SMBH only coevolves with the galaxy bulge as traced by the significant correlation between $\overline{\rm BHAR}$ and $\overline{\rm SFR}$ in bulge-dominated galaxies; while for non-bulge-dominated galaxies, the $\overline{\rm BHAR}$ is not linked with $\overline{\rm SFR}$, but instead, it is predominantly determined by \ms.

However, we note that in \cite{Yang2017} (Y17) and \cite{Yang2019} (Y19), the $\overline{\rm BHAR}$ (calculated from \lxbar~by averaging \lx~for both X-ray-detected and X-ray-undetected galaxies) is derived by assuming a $wabs\times zwabs\times powerlaw$ model, which, as shown in Section 4.1 of paper I, is not appropriate for highly obscured AGNs as it will significantly underestimate \lx~owing to the negligence of the Compton-scattering process. Although their main results will not be influenced by this issue since highly obscured AGNs do not appear to be the dominant population in Y19, and the use of X-ray band in their analysis also minimizes this effect (see Section 3.5.1 of Y17), it is currently unclear whether highly obscured sources follow the same trend as the general AGN population in Y19. Therefore, it is crucial to extend the aforementioned works to the highly obscured regime.

Given the fact that $M_*$ and SFR are positively correlated through the galaxy main-sequence relation and both of them increase with increasing redshifts owning to observational bias or/and actual galaxy evolution, a simple bivariate analysis (e.g., \lx~vs. SFR or \ms) is not able to reveal the leading factor that may predominantly govern the fueling and obscuration environments of SMBH growth. To overcome this issue, we perform  multi-variate linear regression and Spearman partial correlation test using the {\tt{R}} packages {\tt{lm.fit}} and {\tt{ppcor}} of AGN parameters on all three variables simultaneously: \ms, SFR and $z$, which describe how AGN activity and obscuration depend on \ms~(SFR) at given SFR (\ms) and $z$, thereby enabling us to break the degeneracies.  

The linear-regression result for our highly obscured AGN sample using \lx~as a direct tracer of AGN activity is 
\begin{equation}
\begin{aligned}
\loglx & =  (0.21\pm0.06) \times \logms \\ 
& + (0.11\pm0.05) \times \logsfr \\ 
& + (0.28\pm0.04)  \times z + (40.72\pm0.60).
\end{aligned}
\label{eq:lx_dependence}
\end{equation}
There is a statistically significant positive correlation between \lx~and \ms~($3.7 \sigma$)\footnote{The $\sigma$ here represents the significance level that the coefficients deviate from zero.}, and a positive but less significant correlation between \lx~and SFR ($2.4 \sigma$). The regression result is further confirmed by the Spearman partial correlation test that \lx~has a stronger correlation with \ms~($\rho=0.29$ at the $5.1 \sigma$ confidence level) than with SFR ($\rho=0.13$ at the $2.4 \sigma$ confidence level), which appears to be in support of the scenario proposed by Y19 that $\overline{\rm BHAR}$ (equivalent to \lx~in our analysis) is mainly linked with \ms~instead of SFR for non-bulge dominated galaxies (see Section \ref{subsec:morphology} for the result that 71\% of the sources of our morphology sample have a significant disk or irregular morphology). Such enhanced AGN activity in massive galaxies may possibly be related to the greater gravitational potential, which makes it easier to fuel the central SMBH with gas in the vicinity of the nuclear region. Furthermore, we examine whether \lx~for our bulge-dominated galaxies (i.e., the 51 SPHs identified in Section \ref{subsec:morphology}) traces SFR as proposed in Y19. This time we 
do not find any statistically robust correlation between \lx~and \ms~($1.0\sigma$) or SFR ($1.9\sigma$) using Spearman partial correlation tests, perhaps owing to that the small sample size does not properly averaging over all galaxies to assess duty cycle effects and precludes us from finding any significant trend.  

Note that even if \ms~is controlled, there still remains a somewhat weaker positive trend between \lx~and SFR, in agreement with previous studies that reported a positive correlation between \lx~and SFR \citep[e.g.,][]{Lanzuisi2018, Dai2018} and higher average X-ray luminosities in starburst galaxies \citep[e.g.,][]{Rodighiero2015, Grimmett2019}, which could be explained by common cold gas supply for both SF and SMBH accretion. 

We also look for trends of whether AGN obscuration is correlated with host-galaxy properties. The linear regression result for \nh~is
\begin{equation}
\begin{aligned}
\lognh & =  (-0.04\pm0.05) \times \logms \\ 
& +(0.05\pm0.04) \times \logsfr \\ 
& - (0.02\pm0.03) \times z + (24.2\pm0.5),
\end{aligned}
\end{equation}
and the Spearman correlation test yields correlation coefficients consistent with zero for both \ms~($\rho=-0.04$) and SFR ($\rho=0.06$). The lack of correlation is also confirmed by calculating the average \nh~in different $M_*$ or SFR bins, since in either case, we find a flat trend within $1\sigma$ uncertainties \citep[e.g.,][]{Lutz2010, Shao2010, Rosario2012}. The independency of LOS obscuration on host-galaxy properties is naturally expected in the AGN unification model \citep{Antonucci1993}, suggesting that for a significant fraction of our sources, their high \nh~values likely result from high inclination angles, instead of being caused by an intensively dusty environment as a consequence of violent mergers where an enhancement in SFR is expected when the absorbing column density reaches the highly obscured regime \citep[e.g.,][]{Hopkins2008}. The lack of a significant correlation with total stellar mass suggests that for the bulk of X-ray selected highly obscured AGNs, the absorbing materials responsible for the CT-level obscuration are probably confined in the nuclear region.

This finding appears to be inconsistent with some studies in the COSMOS field which reported a somewhat positive \citep[][]{Zou2019} or a strong correlation between obscuration and $M_*$ \citep[][]{Lanzuisi2018}. To alleviate the limitation of the narrow \nh~range being explored, we also include less obscured AGNs while performing partial correlation tests. However, no correlation is detected for either $M_*$ or SFR when the full \nh~range is considered. This discrepancy is likely due to different natures of X-ray and optical obscuration \citep[e.g.,][]{Shimizu2018, Xu2020} as well as sample differences. It is also possible that the narrow $M_*$ range ($\sim 10^{10} - 10^{11} \msun$) of our sample prevents us from finding any significant trend, thus wider surveys with similar depths are required to probe highly obscured AGNs in more massive galaxies and extend our current analyses.

\subsection{Are Highly Obscured AGNs Mainly Triggered by Mergers?}
\label{subsec:morphology}

\begin{figure*}
\centering
\includegraphics[width=\linewidth]{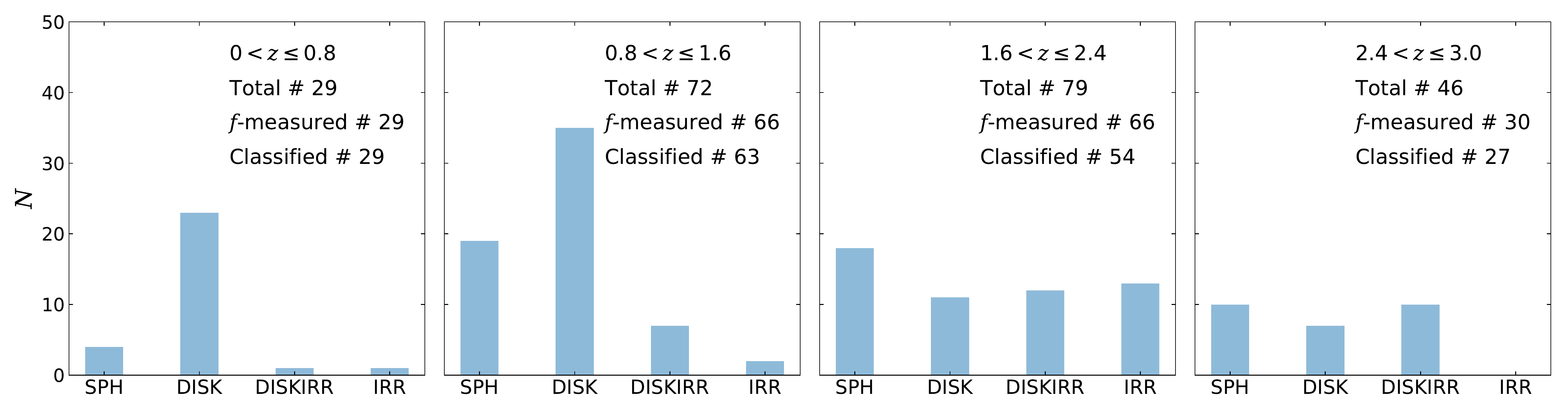}
\caption{Morphological classification results for the 226 highly obscured AGNs matched with the \cite{Company2015} catalog in four redshift bins. In each redshift bin, the total number of matched sources and numbers of sources that have morphological measurements (i.e., $f$-measured) are summarized in the legend. A total of 173 galaxies can be classified into the four classes based on the criteria listed in Section \ref{subsec:morphology}, i.e., 51 SPHs, 76 DISKs, 30 DISKIRRs and 16 IRRs.}
\label{fig:obscured_morphology}
\end{figure*}

In order to investigate the relevance of mergers in triggering highly obscured AGNs, we cross match our sample with the \cite{Company2015} catalog which provides morphology classifications for galaxies with $H$-band magnitude \<~24.5 in the five CANDELS fields based on high-resolution \hst~images and deep-learning techniques. The classification algorithm is trained on the GOODS-S visual-classification results \citep{Kartaltepe2015} and has a very high accuracy. Note that all galaxies in \cite{Company2015} are classified based on $H$-band images, thus we are investigating the rest-frame NIR images for low-redshift sources and rest-frame optical images for high-redshift sources. However, since \cite{Kartaltepe2015} showed that only a small fraction of their sources (i.e., 84 out of 7634 galaxies) have very different classifications  between $V$-band and $H$-band images, we argue that this morphological $k$-correction will not significantly influence our results. 

Since most of our $z>3$ sources do not have measured morphology information in \cite{Company2015}, we exclude them from the morphology analysis. We use the CANDELS counterpart coordinates for our highly obscured AGNs given by the X-ray source catalogs to perform cross-matching. A total of 226 sources are matched using a $0.''5$ matching radius (hereafter the morphology sample). For each galaxy, five parameters are assigned to describe their morphology: $f_{\rm sph}$, $f_{\rm disk}$, $f_{\rm irr}$, $f_{\rm ps}$ and $f_{\rm unc}$, which represent the possibilities that a galaxy is spheroidal, disky, irregular, point-like and unclassifiable, respectively. Among the morphology sample of 226 sources, 191 have a set of the above morphology parameters being derived (hereafter the $f$-measured sources) and we divide them into four groups \citep{Company2015}:

1. {\it pure bulges} (SPH):  $f_{\rm sph} > 2/3$, $f_{\rm disk} < 2/3$ and $f_{\rm irr} < 1/10$;

2. {\it disks} (DISK): $f_{\rm disk} > 2/3$ and $f_{\rm irr} < 1/10$;

3.  {\it irregular disks} (DISKIRR): $f_{\rm disk} > 2/3$, $f_{\rm sph} < 2/3$ and $f_{\rm irr} > 1/10$; and

4.  {\it irregulars/mergers} (IRR): $f_{\rm disk} < 2/3$, $f_{\rm sph} < 2/3$ and $f_{\rm irr} > 1/10$.

Motivated by \cite{Kocevski2015} (see their Section 3), we do not distinguish late-type and early-type disks (i.e., we merge the DISK and DISKSPH groups in \citealt{Company2015} into DISK) to reduce the possible contamination from the AGN to the bulge component. This is also considered for the fact that the disk components are easily destroyed in a major merger event \citep{Hopkins2009}, therefore, as long as a significant undisturbed disk is observable, it is less possible that the galaxies have experienced violent mergers.

Figure \ref{fig:obscured_morphology} presents the distributions of morphology type for our morphology sample in four redshift bins. Among the 191 $f$-measured galaxies, 173 (91\%) of them have been classified as one of the four types, including 51 SPHs, 76 DISKs, 30 DISKIRRs and 16 IRRs. The \zbar~and \loglxbar~of the classified sources are 1.56 and 43.5~\ergs, respectively. Most of these 173 sources ($61\%$) have a significant disk component ($f_{\rm disk} > 2/3$, i.e., DISK + DISKIRR; see also \citealt{Schawinski2012}), while only 27\% of them exhibit irregular signatures (i.e., DISKIRR + IRR).  

For the 18 unclassified sources, 61\% of them (11/18) have $f_{\rm irr}  > 0.1$, with $\overline{f_{\rm irr}} = 0.31$,  $\overline{f_{\rm disk}} = 0.48$ and $\overline{f_{\rm sph}} = 0.76$ (the remaining $f_{\rm irr} \leq 0.1$ sources have $\overline{f_{\rm irr}} = 0.05$,  $\overline{f_{\rm disk}} = 0.47$ and $\overline{f_{\rm sph}} = 0.40$). Visual inspection of their images confirms that some of them do exhibit irregular morphologies, and thus it is possible that these galaxies are experiencing mergers and are transforming their morphology from being disk-dominated to bulge-dominated. If we simply treat all the 11 unclassified $f_{\rm irr} > 0.1$ sources as IRRs, then 57 out of the 191 (30\%) sources show some level of irregularities (i.e., DISKIRR+IRR).  This optimistic fraction is in general agreement with \cite{Kocevski2015} which proposed that $\sim 22\%$ of X-ray-selected highly obscured AGNs exhibit merger or interaction signatures.

Such a small irregular fraction suggests that major mergers cannot be the leading mechanism that triggers highly obscured SMBH accretion, especially considering the fact that an irregular disk morphology does not necessarily implies galaxy interactions.\footnote{For example, the DISKIRR morphology could be a result from minor mergers, the extended signatures from major mergers that are misclassified as disks; or it could be due to strong disk instabilities in high-redshift gas-rich galaxies.} The majority (61\%) of the classified sources having a significant disk component (i.e., 76 DISKs + 30 DISKIRRs) can be considered as a further argument against the major-merger scenario, which disfavors the probability that the likely time lag between the merger and the later onset of nuclear activity \citep[e.g.,][]{Emonts2006, Hopkins2012} prevents us from finding merger signatures that have already faded, as it is not very likely for the disk structure to survive after experiencing the violent merger process for ordinary galaxies \citep[e.g.,][]{Hopkins2009}. 

\begin{table*}
\centering
\caption{DISKIRR and IRR fractions for the combined highly obscured and less obscured AGN samples in different \nh~and \lx~bins.}
\begin{tabular}{c c c c c c c}
\hline
\hline
Fraction & Total & $\nh < \cm$ & $\cm < \nh < \ct$ & $\nh > \ct$ & $\lx < 10^{44}\ \ergs$ & $\lx >10^{44}\ \ergs$\\
\hline
$f_{\rm DISKIRR}$ & $15_{-2}^{+2}\%$ & $12_{-2}^{+3}\%$ & $16_{-3}^{+4}\%$ & $22_{-5}^{+7}\%$ & $13_{-2}^{+2}\%$ & $26_{-5}^{+7}$\%\\ 

$f_{\rm IRR}$ & $11_{-1}^{+2}\%$ & $8_{-2}^{+3}\%$ & $8_{-2}^{+3}\%$ & $12_{-3}^{+6}\%$ & $10_{-2}^{+2}\%$ & $11_{-3}^{+6}\%$\\
\hline

\end{tabular}
\label{table:irr_frac}
\end{table*}

\begin{figure*}
\centering
\includegraphics[width=0.32\linewidth]{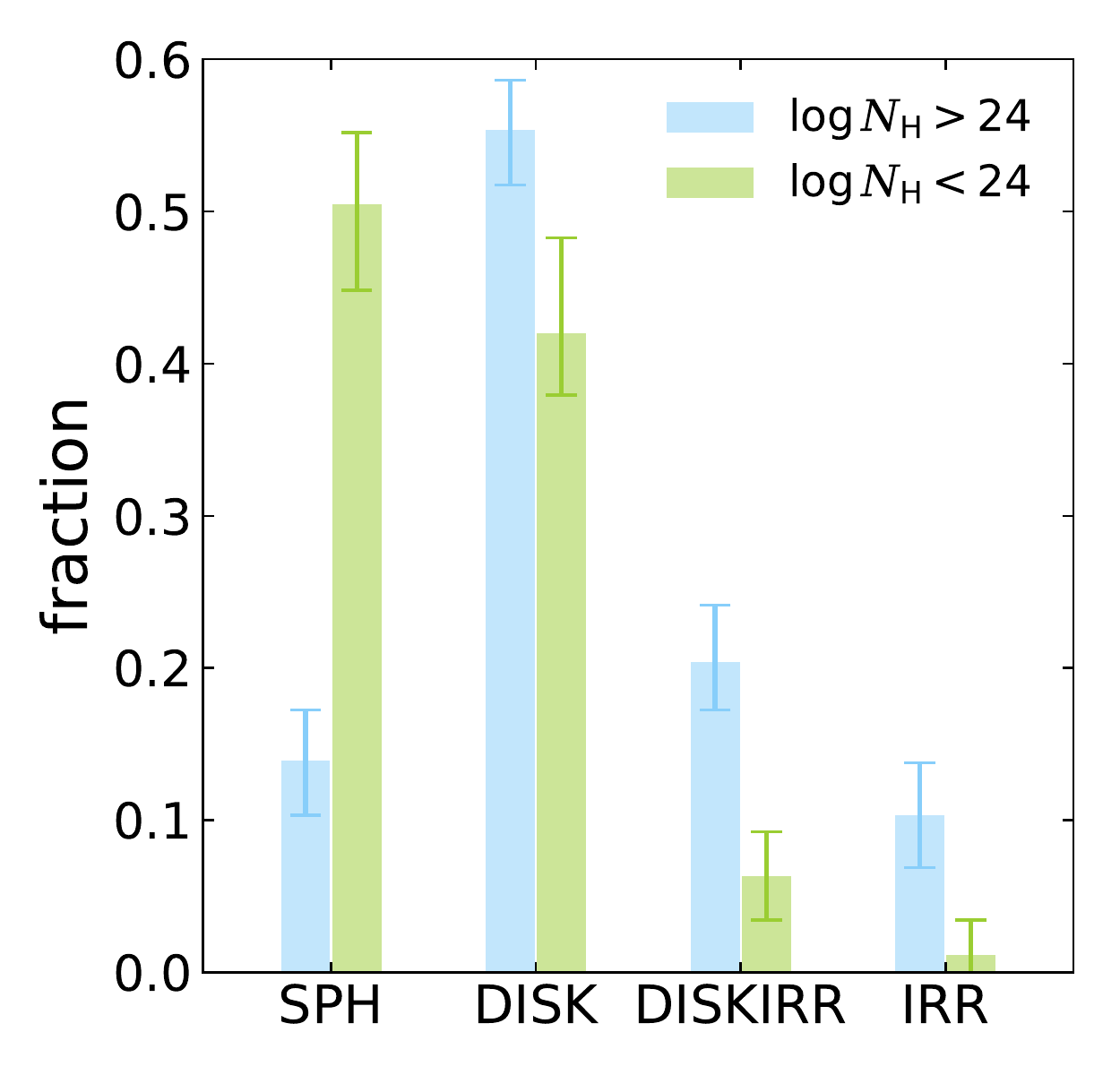}
\includegraphics[width=0.32\linewidth]{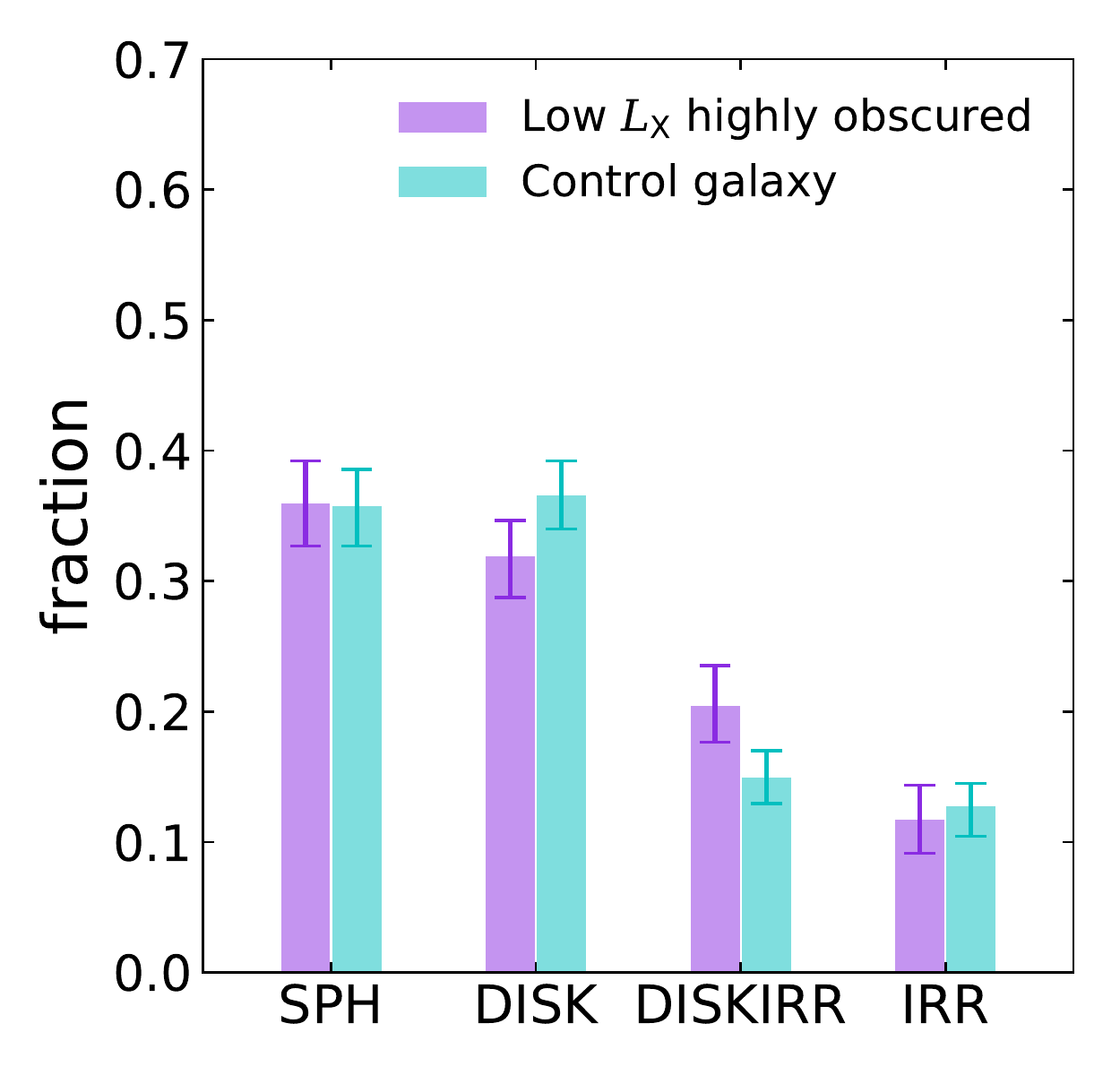}
\includegraphics[width=0.32\linewidth]{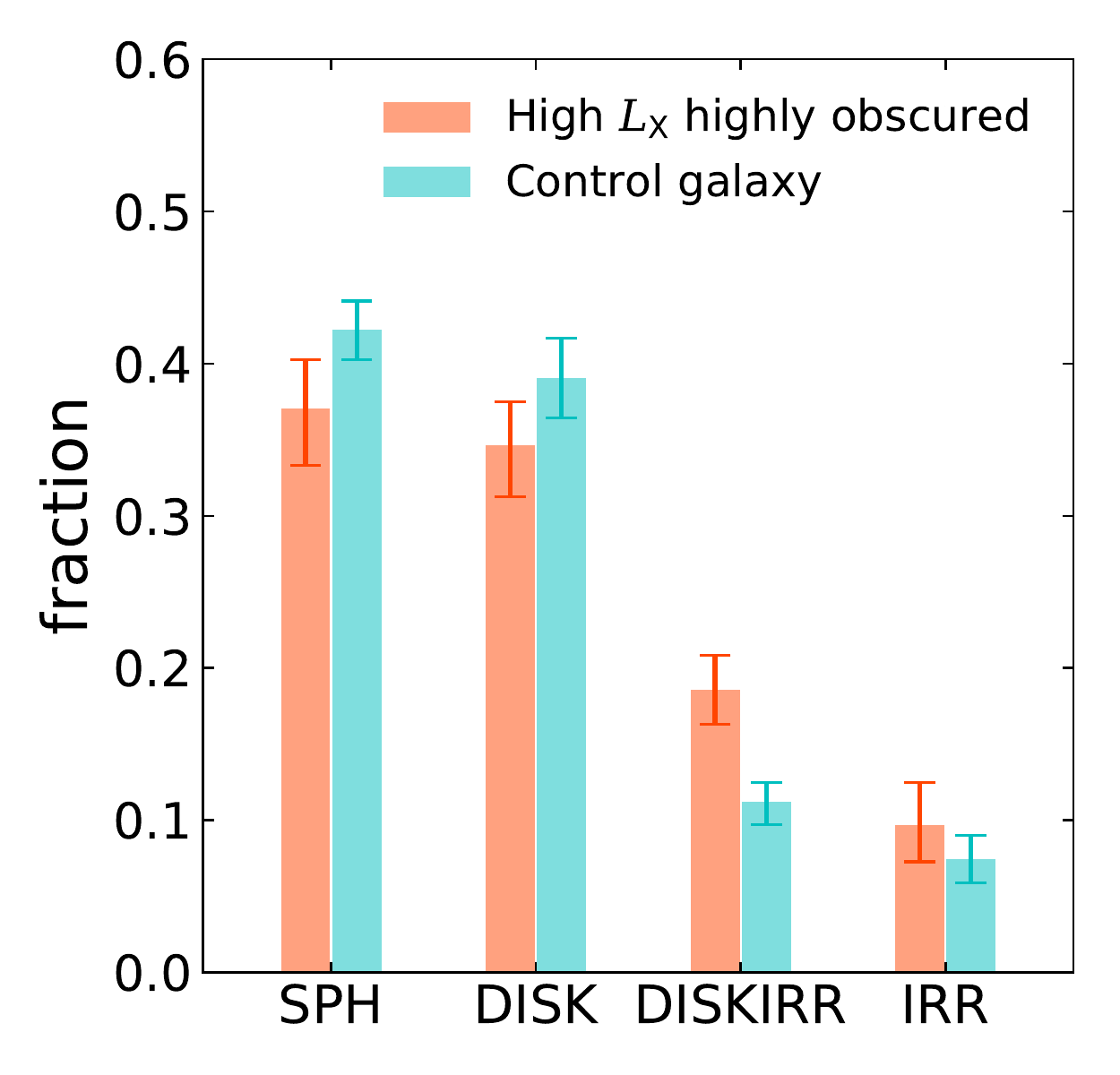}
\caption{Comparison of morphology-classification distributions between \lx-~and $z$-matched CT and CN AGNs (Left), between \ms-~and $z$-matched low-\lx~(i.e., \lx~smaller than the median X-ray luminosity value at each redshift grid) highly obscured AGNs and non-active galaxies (Middle), and between \ms-~and $z$-matched high-\lx~(i.e., \lx~larger than the median X-ray luminosity value at each redshift grid) highly obscured AGNs and non-active galaxies (Right), respectively. The $1\sigma$ error bars are derived from the 16th- and 84th-percentiles of the 1000 resampled distributions.}
\label{fig:control_morphology}
\end{figure*}

However, it has been argued that major mergers may only be important in triggering the most-luminous  AGNs and/or the most obscured CT AGNs \citep[e.g.,][]{Treister2012, Kocevski2015, Chang2017b}.  To examine such arguments, we perform a similar morphological analysis on less obscured AGNs and show the DISKIRR and IRR fractions in different \nh~and \lx~bins In Table \ref{table:irr_frac}. The $1\sigma$ errors are calculated using the method of \cite{Cameron2011}. We  confirm the trend presented in \cite{Kocevski2015} that the irregular fraction increases with \nh. However, we note that the average \lx~is 0.8 dex higher for the CT population, and an elevated irregular fraction in high-\lx~bin is also detected. 
To make a fair comparison, we construct \lx-~and $z$-controlled CT and CN AGN subsamples using the same method described in Section \ref{subsec:msq}, and the result is plotted in the left panel of Figure \ref{fig:control_morphology}. It can be seen that, although the sample is very limited (only 29 CT AGNs have $\lx$- and $z$-matched CN counterparts with $\loglxbar = 43.7$,  thus we do not further divide the CN subsample into different \nh~bins), both DISKIRR and IRR fractions are  higher in CT AGNs than their low-\nh~counterparts, suggesting that the observed difference in \nh~should not be only caused by inclination effects. The elevated irregular fraction in the most heavily obscured CT population implies that mergers indeed play a part in boosting the nuclear obscuration \citep[e.g.,][]{Ricci2017merg, Koss2018}. 

However, we notice that in either population (high \nh~or high \lx), the DISKIRR and IRR classes still only occupy a small fraction of the total sample. To better understand the importance of mergers in triggering highly obscured AGNs in the context of galaxy evolution, we split our morphology sample into two subsamples based on the median \lx~in each redshift grid and construct the \ms-~and $z$-controlled normal galaxy sample for each population from the S15 GOODS-S dataset. Their morphology-classification results are then compared with our highly obscured AGNs using the same classification criteria. The middle and right panels of  Figure \ref{fig:control_morphology} show the average morphology-classification distributions for the 1000 randomly-selected control-galaxy samples and our highly obscured AGNs in the low-\lx~and high-\lx~regimes, respectively. 
For the low-luminosity highly obscured population, the irregular fraction is indistinguishable from that of control galaxies in terms of both DISKIRR fraction ($20_{-3}^{+3}\%$ vs. $16_{-3}^{+3}\%$) and IRR fraction ($12_{-3}^{+3}\%$ vs. $13_{-3}^{+3}\%$); while for the high-luminosity highly obscured population, the IRR fraction ($8_{-2}^{+2}\%$) is similar to that of control galaxies ($9_{-3}^{+3}\%$), but the DISKIRR fraction increases from $12_{-3}^{+3}\%$ for control galaxies to $19_{-3}^{+3}\%$ for luminous highly obscured AGNs. This, together with the result that the irregular fraction increases with \nh, suggests that galaxy interactions are more relevant in triggering the most-luminous and the most heavily obscured (X-ray-selected) AGNs; however, it may still play a limited role as even for such extreme populations, the disturbed hosts are still in the minority.

Therefore, although mergers do have the ability to trigger highly obscured AGNs \citep[e.g.,][]{Ricci2017merg, Goulding2018, DeRosa2018, Pfeifle2019}, our result here argues that the statement that the majority of X-ray-selected highly obscured AGNs are triggered by mergers is not true. 
The large fractions of undisturbed disk and spheroid hosts for our highly obscured sample shown in Figure~\ref{fig:control_morphology}  implies that secular processes (e.g., galactic disk instabilities) should be the predominant triggering mechanism of the current highly obscured SMBH accretion by fueling cold gas streams to the central AGNs stochastically \citep[e.g.,][]{Schawinski2012, Kocevski2012, Kocevski2015, Chang2017a}

\subsection{Are Highly Obscured AGNs Experiencing a Blow-out Phase?}\label{sec:outflow}

\begin{figure}
\centering
\includegraphics[width=\linewidth]{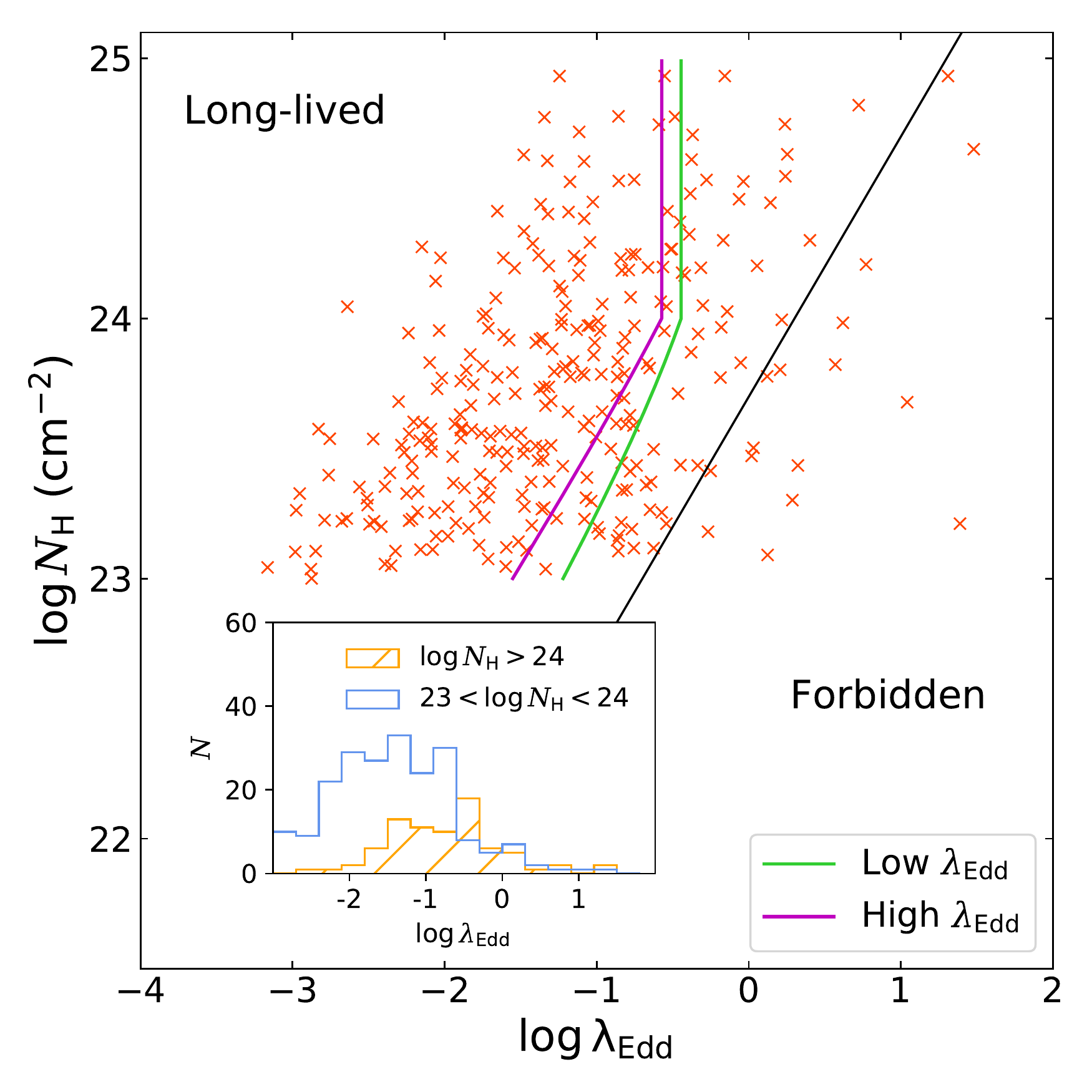}
\caption{\nh~vs. \edd~plane for highly obscured AGNs. The typical uncertainty in \edd~is estimated to be $\sim 0.8$ dex. The long-lived and forbidden regions defined in \cite{Fabian2008} are separated by the black line. 
The green and magenta curves (derived by assuming typical SEDs for low- and high-\edd~AGNs while calculating the boost factor $A$, respectively) represent the minimum \edd~values that are required to blow out the surrounding dusty materials in the \cite{Liu2011} anisotropic radiative feedback model. Inset: Distributions of \logedd~for CT and highly obscured CN AGNs.}
\label{fig:outflow}
\end{figure}

Blueshifted absorption lines detected in AGNs have been widely interpreted as an indicator of outflowing materials \citep[e.g.,][]{Gibson2009, FilizAk2013, Li2019b}, which have long been served as crucial ingredients in the evolutionary models that are responsible for transforming AGN types (i.e., from type 2 to type 1 or X-ray obscured to unobscured) and quenching star formation (see, e.g., \citealt{King2015} for a review). 
A possible way to study whether our highly obscured AGNs are experiencing a blow-out phase and may eventually become unobscured AGNs is to  investigate whether they are in the ``forbidden'' outflow region in the \nh~vs. \edd~diagram as defined in \cite{Fabian2008}.
To obtain \edd, X-ray luminosity is converted into bolometric luminosity (\lbol) using the \cite{Lusso2012} luminosity-dependent conversion factor in the form of 
\begin{equation}
{\rm log}\, L_{\rm bol}\,/\,L_{\rm X} = 0.230\,x + 0.050\, x^2 + 0.001 \,x^3 + 1.256,
\end{equation}
where  $x = {\rm log}\, L_{\rm bol} - 12$. The black-hole mass is obtained by scaling the total stellar mass using the relation proposed by 
\cite{Sun2015} that is parameterized as ${\rm log}\,M_{\rm BH}/\ms = -2.85$.
The \mbh~calculated from this method will inevitably suffer large uncertainties, but we note that it is still possible to obtain the average \mbh~information for our sample, as various studies have observed a positive correlation between \mbh~and \ms, albeit with large dispersions \citep[e.g.,][]{Merloni2010, Sun2015, Reines2015}.  The \edd~is then calculated as $\lbol / \ledd$ where $\ledd = 1.26 \times 10^{38}\ (M_{\rm BH}\,/\,\msun)$ erg~s$^{-1}$. The typical uncertainty in \edd~is $\sim0.8$~dex, which is estimated by combining the typical uncertainties (i.e., the median value of the whole sample) from the \lx~($\sim 0.25$~dex) and the $M_*$ ($\sim 0.05$~dex) measurements,  as well as the scatters of the $\lx - \lbol$ ($\sim 0.2$~dex; \citealt{Lusso2012}) and $\mbh - M_*$ ($\sim 0.3$~dex; \citealt{Sun2015}) relations.

The \nh~vs. \edd~relation is plotted in Figure~\ref{fig:outflow}, where the distribution of \edd~is also shown. It can be seen that the bulk of our sources are now in the long-lived region \citep[e.g.,][]{Raimundo2010, Ricci2017nat}. As pointed out by \cite{Liu2011}, for the effective Eddington ratio $\lambda = A \edd$ below a critical value of 7/18, where $A$ is the boost factor of radiation pressure owing to the presence of dust, the gravitational force will always defeat radiation pressure in all directions, hence the dusty materials will not be blown out. Adopting the boost factor $A$ calculated by \cite{Fabian2008}, we derive the critical \edd~that a radiatively-driven outflow can be launched as a function of \nh~and display the results in Figure \ref{fig:outflow}. The majority of our sources are located in the left side of the critical curves, suggesting that their current obscuring materials are long-lived in all directions given their instantaneous \edd. 
Given that these sources have smaller $\overline{\rm log\,\edd}$~(--1.5), $\rm \overline{log\,sSFR}$ (--9.6/yr) and larger $\overline{\logms}$~(10.7~\msun) than those sources on the right side of the critical curves with the corresponding values of --0.2, --8.6/yr and 10.2~\msun, respectively,
they may have already evolved to a less active phase in terms of both star formation and BHAR, thus their obscuring materials are likely to survive from the  feedback events and they will remain X-ray  obscured, instead of transforming to X-ray unobscured AGNs.
However, simulations have shown that \edd~can vary by orders of magnitude on timescales of $\sim 10^5$-$10^6$ yrs \citep[e.g.,][]{Novak2011, Yuan2018}. Therefore, we cannot rule out the possibility that once a new cycle of significant SMBH accretion is being triggered (e.g., through mergers), these obscuring materials may be cleared out and an unobscured AGN may be revealed along our LOS.

\section{Conclusions and Discussions}
\label{sec:conclusions}
In this work, by using a large sample of X-ray-selected $\nh > \cm$ AGNs (294 sources at $z = 0-5$) and the wealth of spectroscopic and photometric data available in CDFs, we have carried out a systematic multiwavelength study of highly obscured AGNs that is supplementary to our previous X-ray spectral and long-term variability analyses (see paper~I), aiming at examining whether highly obscured AGNs are the missing link in the merger-triggered SMBH-galaxy coevolutionary models.

Specifically, we investigate the distributions of our X-ray sources on various optical/IR/X-ray color-color diagrams to explore the AGN obscuration properties. We also perform detailed multi-component SED decomposition using the X-CIGALE code \citep{Yang2020} to obtain crucial host-galaxy parameters (e.g., \ms~and SFR). The inclusion of X-ray data and the use of the clumpy torus model in SED fitting allow us to better constrain the AGN power and thereby obtaining more reliable host-galaxy properties.
We explore potential correlations among $M_*$ and SFR with the direct tracers of AGN radiation power and obscuration, i.e., the X-ray luminosity and column density derived from X-ray spectral fitting, in order to search for possible connections between the growths of SMBHs and their host galaxies. The morphology classification is then performed for the purpose of evaluating the importance of mergers in triggering highly obscured SMBH accretion. Lastly, we present our analysis on whether highly obscured AGNs are sweeping out the surrounding obscuring materials which may ultimately make them unobscured AGNs.
The primary conclusions emerging from this work are summarized as follows.

\begin{enumerate}[1.]

\item The \cite{Donley2012} IRAC color-color diagram can successfully identify a substantial fraction of highly obscured AGNs at $\loglx\, (\ergs) > 44.5$. However, the identification fraction dramatically drops to $\lesssim 20\%$ even for sources with $44.0 < \loglx\, (\ergs) < 44.5$. The low identification efficiency for X-ray luminous ($\loglx > 44.0\ \ergs$) highly obscured AGNs is likely due to that the IRAC-missed sources have lower dust contents and/or torus CFs, and are therefore intrinsically fainter in MIR ($\loglmirbar = 44.1\ \ergs$) than the IRAC-selected ones ($\loglmirbar = 45.1\ \ergs$), given that the average X-ray luminosities for the two populations are the same ($\loglxbar = 44.4\ \ergs$; see Section \ref{subsec:irac}). 

\item A large fraction (84\%) of our X-ray  highly obscured AGNs will not be selected as heavily obscured candidates using the flux ratio of $\fmirr > 1000$ and $R-K > 4.5$ criteria proposed by \cite{Fiore2008}, even for the most-luminous (i.e., $\lx > 10^{44}\rm \ erg\ s^{-1}$) population (i.e., 70\% being missed). 
This result suggests that the heaviest X-ray obscuration is not equivalent to the extremely large MIR-to-optical flux ratios and the reddest colors, possibly owing to the diverse structures of obscuring materials (e.g., different CFs, gas/dust contents, and \nh~distribution), complex origins of the LOS X-ray obscuration (e.g., dust-free BLR gas, dusty torus, disk wind and/or ISM) as well as galaxy contamination to the observed colors (see Section \ref{subsec:f24}).

\item The 50th-percentile of the SFR distribution for highly obscured AGN hosts is similar to that of \ms-~and $z$-controlled normal galaxies ($\overline{\Delta {\rm log\,SFR}_{\rm 50th,\, agn-galaxy}}= -0.03_{-0.12}^{+0.12}$), but the 20th-percentile for the former is characteristically larger than that of the latter ($\overline{\Delta {\rm log\,SFR}_{\rm 20th\,, agn-galaxy}}= 0.83_{-0.20}^{+0.17}$). As a result, the SFR distribution of highly obscured AGNs is narrower than normal galaxies (i.e., lack of quiescent hosts), suggesting that the sufficient cold gas supply that is responsible for maintaining star formation may also be an important source in fueling highly obscured SMBH accretion. Furthermore, the 80th-percentile of the SFR distribution of highly obscured AGNs is not enhanced relative to control galaxies ($\overline{\Delta {\rm log\,SFR}_{\rm 80th,\, agn-galaxy}}= -0.30_{-0.07}^{+0.08}$), suggesting that the presence of X-ray-selected highly obscured AGNs is not more frequently connected to starburst activities (see Section \ref{subsec:msq}).

\item The multi-variate linear regression and Spearman partial correlation analyses among \lx, \ms, SFR and $z$ suggest that highly obscured SMBH accretion (traced by \lx) is more fundamentally related to \ms, which might be a result of the greater gravitational potentials of massive hosts, being consistent with previous studies on non-bulge-dominated galaxies (see conclusion 6). However, highly obscured SMBH accretion still remains a positive trend with SFR after $M_*$ is controlled, albeit at a weaker significance level, suggesting that at a given \ms, galaxies with sufficient gas contents which are able to fuel higher SFRs are also more likely to trigger highly obscured AGNs (see Section \ref{subsec:link}).  

\item We find no correlation between \nh~and either \ms~or SFR, consistent with the prediction from the AGN unification model \citep{Antonucci1993}. Such a result suggests that for a significant fraction of our sources, their high \nh~values likely result from high inclination angles, instead of being caused by an intensively dusty environment as a consequence of violent mergers where an enhancement in SFR is expected when the absorbing column densities reach the highly obscured regime \citep[e.g.,][]{Hopkins2008}. The lack of a significant correlation with total stellar mass suggests that for the bulk of X-ray selected highly obscured AGNs, the absorbing materials responsible for the CT-level obscuration are probably confined in the nuclear region.

\item To examine whether highly obscured SMBH accretion is mainly triggered by mergers, we cross-match our sample with the \cite{Company2015} galaxy-morphology catalog. We find that $61\%$ of them have a significant disk component, while only $27\%$ of them exhibit irregular signatures (9\% IRR + 18\% DISKIRR). The incidence of disturbed morphologies increasing with both \lx~and \nh, which supports the scenario that mergers are more relevant in triggering the most-luminous and the most heavily obscured (X-ray-selected) AGNs. However, mergers may still only play a limited role as even for such extreme populations, the disturbed hosts are still in the minority (see Section \ref{subsec:morphology}). 

\item The majority of our sources are located in the stable long-lived region in the \nh~vs. \edd~plane defined in \cite{Fabian2008}. The fact that long-lived sources have on average smaller \eddbar~and \ssfrbar, and higher \mbar~with respect to sources that have matched $\overline{N_{\rm H,\, LOS}}$~but lie in or close to the outflow region suggests that, 
they may have already evolved to a less active phase in terms of both star formation and BHAR, thus their obscuring materials are likely to survive from the feedback events and they will remain X-ray obscured. However, we cannot rule out the possibility that once a new cycle of significant SMBH accretion activity is being triggered, these obscuring materials may be cleared out and an unobscured AGN will be revealed along our LOS (see Section \ref{sec:outflow}).
\end{enumerate}

Overall, our findings suggest that the majority of the X-ray-selected highly obscured AGNs within the luminosity and redshift ranges examined here are not the ``missing link'' (i.e., dust-enshrouded AGNs likely having enhanced star-forming activities, disturbed host-galaxy morphologies, and strong outflows which may eventually make themselves unobscured) in the merger-triggered SMBH-galaxy coevolution and AGN type-transition models, given the combined evidence of complex origins of their heavy LOS X-ray obscuration, the similar star formation activity to that of normal star-forming galaxies, the lack of correlation between absorbing column densities and SFR, the large fraction of sources having undisturbed hosts and the fact that the majority of them are far away from the outflow region defined in the \nh~vs. \edd~plane.

However, although our work has provided deep insights into the elusive highly obscured AGN population by taking advantage of the deepest \chandra~surveys, the current X-ray data are still insufficient. The small sky coverages of CDFs severely limit the source statistics toward the highest-luminosity end, and the current detection limit for \chandra~is insufficient to probe CT AGNs that have \nh~as high as $10^{25}\ \nhu$. Therefore, we are still missing the most luminous and/or the most heavily obscured AGNs which may be most important to understand the merger models (see Section \ref{subsec:morphology}). To improve the current situation, wider X-ray surveys with similar or  deeper depth are necessary. This requirement is expected to be fulfilled by future {\it{Athena}} \citep{Nandra2013} and {\it{Lynx}} \citep{Gaskin2019} surveys which can reach detection limits close to (i.e.,  {\it{Athena}}) or even deeper (i.e., {\it{Lynx}}) than the current \chandra~deep surveys while also achieving much larger survey areas. In concert with the improved constraints of dust properties and star formation and AGN activities at IR wavelengths by future MIR spectroscopy from e.g., the {\it{James Webb Space Telescope}} \citep[e.g.,][]{Kirkpatrick2017}  and the {\it{SPICA}} mission \citep[e.g.,][]{spica}, we will be able to better address the issues related to current studies and have a more profound understanding of the highly obscured AGN population as well as their role in galaxy evolution.

\acknowledgments 
We thank the referee for helpful suggestions. J.Y.L. and Y.Q.X. acknowledge support from the National Natural Science Foundation of China (NSFC-11890693, 11421303), the CAS Frontier Science Key Research Program (QYZDJ-SSW-SLH006), and the K.C. Wong Education Foundation. 
M.Y.S acknowledges support from the National Natural Science Foundation
of China (NSFC-11973002).
W.N.B. acknowledges support from Chandra X-ray Center grant GO8-19076X, NASA grant 80NSSC19K0961, and Penn State ACIS Instrument Team Contract SV4-74018 (issued by the Chandra X-ray Center, which is operated by the Smithsonian Astrophysical Observatory for and on behalf of NASA under contract NAS8-03060). The Chandra ACIS team Guaranteed Time Observations (GTO) utilized were selected by the ACIS Instrument Principal Investigator, Gordon P. Garmire, currently of the Huntingdon Institute for X-ray Astronomy, LLC, which is under contract to the Smithsonian Astrophysical Observatory via Contract SV2-82024.
F.V. acknowledges financial support from CONICYT and CASSACA through the Fourth call for tenders of the CAS-CONICYT Fund, and CONICYT grants Basal-CATA AFB-170002. 
P.T. acknowledges financial contribution from the agreement ASI-INAF n.2017-14-H.0.
L.L.F. acknowledges the support from the National Natural Science Foundation of China (NSFC-11822303 and 11773020) and Shandong Provincial Natural Science Foundation, China (ZR2017QA001, JQ201801).
 

\end{document}